\documentclass[12pt]{elsarticle}
\usepackage[dvipsnames]{xcolor}
\RequirePackage[T1]{fontenc}
\RequirePackage{mathptmx}
\usepackage{fullpage}
\usepackage{slashed}
\usepackage{graphics,bm}
\usepackage{epsfig}
\usepackage{graphicx}
\usepackage{amsmath}
\usepackage{amssymb}
\usepackage{bbold}
\usepackage{wrapfig}
\usepackage{color}
\usepackage{hyperref}
\usepackage{mathtools}
\usepackage{cancel}
\usepackage{multirow}
\usepackage{upgreek}
\usepackage{subcaption}
\usepackage{feynmf}
    \usepackage[vertfit]{breakurl} 

\journal{}
\DeclareMathOperator{\Tr}{Tr}
\usepackage{inputenc}
\usepackage{physics}
\usepackage{bbm}
\usepackage[compat=1.1.0]{tikz-feynman}
\usepackage{soul}
\newcommand{\beq}{\begin{equation}\begin{aligned}}
\newcommand{\eeq}{\end{aligned}\end{equation}}
\newcommand{\bqa}{\begin{eqnarray}}
\newcommand{\eqa}{\end{eqnarray}}
\usepackage{xcolor,cancel}
\usepackage[export]{adjustbox}
\usepackage[final]{pdfpages}
\usepackage{pgfplots}
\usepackage{placeins}

\pgfplotsset{compat=1.9}
\pgfplotsset{
    tick label style={font=\footnotesize},
    label style={font=\footnotesize},
    legend style={font=\tiny},
    every axis plot/.append style={line width=0.5pt},
    }
\begin{document}
\def\a{{\alpha}}
\def\be{{\beta}}
\def\d{{\delta}}
\def\D{{\Delta}}
\def\P{{\Pi}}
\def\p{{\pi}}
\def\e{{\varepsilon}}
\def\ep{{\epsilon}}
\def\g{{\gamma}}
\def\k{{\kappa}}
\def\l{{\lambda}}
\def\L{{\Lambda}}
\def\m{{\mu}}
\def\n{{\nu}}
\def\o{{\omega}}
\def\O{{\Omega}}
\def\S{{\Sigma}}
\def\s{{\sigma}}
\def\t{{\tau}}
\def\x{{\xi}}
\def\X{{\Xi}}
\def\z{{\zeta}}

\def\ol#1{{\overline{#1}}}
\def\c#1{{\mathcal{#1}}}
\def\b#1{{\bm{#1}}}
\def\eqref#1{{(\ref{#1})}}

\def\ed#1{{\textcolor{magenta}{#1}}}
\def\edd#1{{\textcolor{cyan}{#1}}}
\begin{frontmatter}
\title{Vacuum free energy, quark condensate shifts and magnetization in three-flavor chiral perturbation theory to $\mathcal{O}(p^6)$  in a uniform magnetic field}

\author{Prabal Adhikari~\footnotemark[1]\footnotetext[1]{The author is currently also a KITP Scholar at the Kavli Institute for Theoretical Physics, University of California, Santa Barbara, CA 93106.}}
\ead{adhika1@stolaf.edu}
\address{Physics Department, Faculty of Natural Sciences and Mathematics, St. Olaf College, 1520 St. Olaf Avenue, Northfield, MN 55057, United States}
\author{Inga Str{\"u}mke}
\ead{inga.strumke@ntnu.no}
\address{Department of Computer Science, Faculty of Information Technology and Electrical Engineering, NTNU, Norwegian University of Science and Technology, Sem S{\ae}landsvei 9, N-7491 Trondheim, Norway}
\date{\today}
\begin{abstract}
We study three-flavor QCD in a uniform magnetic field using chiral perturbation theory ($\chi$PT). We construct the vacuum free energy density, quark condensate shifts induced by the magnetic field and the renormalized magnetization to $\mathcal{O}(p^6)$ in the chiral expansion. We find that the calculation of the free energy is greatly simplified by cancellations among two-loop diagrams involving charged mesons. In comparing our results with recent $2+1$-flavor lattice QCD data, we find that the light quark condensate shift at $\mathcal{O}(p^6)$ is in better agreement than the shift at $\mathcal{O}(p^4)$. We also find that the renormalized magnetization, due to its smallness, possesses large uncertainties at $\mathcal{O}(p^{6})$ due to the uncertainties in the low-energy constants.
\end{abstract}
\end{frontmatter}
\section{Introduction}
\label{sec:introduction}
Quantum Chromodynamics (QCD) in a magnetic background has generated interest in recent years, due to its phenomenological importance to the astrophysics of neutron stars, which are in a cold and possibly magnetized state (within magnetars),  and their relevance in non-central heavy-ion collisions in which QCD may undergo a transition to a deconfined, high temperature phase~\cite{kharzeev2013strongly,Andersen:2014xxa,bandyopadhyay2021inverse,andersen2021qcd}. Furthermore, the QCD phase diagram at finite magnetic field has generated interest even at zero temperature and zero 
baryon density due to its roles in modifying the chiral condensate, the chiral order parameter that characterizes the ground state of QCD. For two massless flavors, the QCD Lagrangian has an $SU(2)_{L}\times SU(2)_{R}\times U(1)_{B}$ symmetry associated with independent rotations of the left and right handed quarks, the ground state breaks this symmetry down to an $SU(2)_{V}\times U(1)_{B}$ giving rise to three Goldstone modes, i.e.\ pions in the low-energy spectrum.  For three massless flavors, the Lagrangian has an $SU(3)_{L}\times SU(3)_{R}\times U(1)_{B}$ symmetry which is broken down to $SU(3)_{V}$ by the formation of the light and strange quark condensates giving rise to eight Goldstone modes, namely the three pions, four kaons and an eta, consistent with Goldstone's theorem. 
In reality these degrees of freedom have masses that are smaller than the typical hadronic scale of approximately $\Lambda_{\rm Had}\sim1\ {\tt GeV}$ due to the finite constituent quark masses and are referred to as pseudo-Goldstone modes.
The effective field theory that encapsulates the interactions of these low-energy degrees of freedom is chiral perturbation theory ($\chi$PT)~\cite{Weinberg:1978kz,Gasser:1983yg,Gasser:1984gg,bijnens1999mesonic,bijnens2000renormalization}.
$\chi$PT is constructed solely using the global symmetries of QCD and its low-energy degrees of freedom. Given a consistent power-counting scheme, one can systematically calculate model-independent corrections to processes involving pions and kaons in a low-energy expansion.

While there are numerous model-dependent studies of the QCD vacuum in the presence of electromagnetic fields, here we focus on $\chi$PT, which is a model-independent low-energy effective theory of QCD.  Studies at  $\mathcal{O}(p^4)$ of two-flavor $\chi$PT in a uniform magnetic field at $T=0$ were conducted in Refs.~\cite{shushpanov1997quark,Cohen:2007bt} using the Schwinger formalism~\cite{Schwinger:1951nm} first developed in the context of quantum electrodynamics. The standard Schwinger integral for the effective potential of charged bosons gives rise to a magnetic field-dependent contribution that decreases in magnitude with increasing boson mass and increases with increasing magnetic fields. The chiral condensate, which is negative in the QCD vacuum, is in effect a measure of the first-order change of the vacuum energy as a function of the quark mass, and increases in magnitude with increasing magnetic fields. This is an example of magnetic catalysis, first discussed in Refs.~\cite{klevansky1989chiral,suganuma1991behavior,klimenko1992three,Klimenko:1990rh,Klimenko:1992ch,gusynin1994catalysis,gusynin1996dimensional}. At $T=0$,  magnetic catalysis is a robust phenomenon observed in low-energy models and theories as well on the lattice~\cite{buividovich2010numerical,buividovich2010chiral,braguta2012chiral,d2011chiral,bali2012qcd,ding2020chiral}. On the lattice, the mechanism behind magnetic catalysis can be understood in terms of the so-called valence and sea contributions, as discussed  in e.g.\ Ref.~\cite{d2011chiral}. To a very good approximation, the change of the quark condensate with increasing magnetic field is the sum of terms coming separately from the change in the Dirac operator and the measure in the partition function. At zero temperature, both contributions enhance the condensate as the field increases. On the other hand, inverse magnetic catalysis refers to either to the decrease in the deconfinement transition temperature upon the introduction of an external magnetic field, or to the decrease in the size of the chiral condensate with increasing temperature as observed in lattice QCD calculations. As discussed in Refs.~\cite{andersen2012thermal,andersen2012chiral}, $\chi$PT calculations at $\c{O}(p^{6})$ produce a behavior that is opposite to those observed in lattice calculations.

Zero temperature studies at $\mathcal{O}(p^6)$ in two-flavor $\chi$PT were first conducted in Refs.~\cite{agasian2000quark,Werbos:2007ym}. In Ref.~\cite{Werbos:2007ym}, the conclusion is that due to the uncertainties in the low-energy constants of $\chi$PT, the chiral condensate may or may not be enhanced in magnitude relative to the  $\mathcal{O}(p^4)$ values. In this paper, we study in addition to the chiral condensate and magnetic catalysis, the renormalized magnetization of the QCD vacuum in a uniform magnetic background in three-flavor $\chi$PT to $\mathcal{O}(p^6)$.
Magnetization, the response of the vacuum free energy to a first order change in the magnetic field or the derivative of the vacuum free energy with respect to the external magnetic field, was first considered briefly within $\chi$PT in Ref.~\cite{kabat2002qcd}, and more recently in Ref.~\cite{hofmann2021diamagnetic}, in the deconfined phase of QCD in Ref.~\cite{cohen2009magnetization}, and the hadron resonance gas model in Ref.~\cite{endrHodi2013qcd}. Within lattice QCD, it has been studied in Ref.~\cite{bali2013magnetic}.

In this paper, we generalize the $\c{O}(p^{4})$ $\chi$PT comparison to lattice QCD of Ref.~\cite{Adhikari:2021bou} and calculate the vacuum free energy density, the condensate shifts, and the renormalized magnetization in three-flavor $\chi$PT to $\mathcal{O}(p^6)$. The paper is organized as follows. In Section~\ref{sec:Lagrangian}, we discuss the relevant $\chi$PT Lagrangian required for the calculation of the free energy (density). In the following Section~\ref{sec:propagator}, we derive the translationally non-invariant, Schwinger propagator associated with mesons in a background magnetic field and discuss simplifying features of the two-loop mesonic diagrams. In Section~\ref{sec:thermodynamics}, we discuss the free energy, quark condensates and magnetization at zero temperature, compare with recent lattice calculations and conclude with a summary in Section~\ref{sec:summary}. Finally, we list the various constants and renormalized $\chi$PT low energy constants required for renormalization in \ref{app:usefulLEC}, the relevant $\chi$PT Lagrangian in \ref{app:L}, useful Schwinger integrals in \ref{app:integrals} and the zero magnetic field vacuum free energy in terms of bare quantities in \ref{app:freeenergy}.

\section{The $\chi$PT Lagrangian}
\label{sec:Lagrangian}
The fundamental building blocks of the $\chi$PT Lagrangian are the pseudo-Goldstone modes, which are encoded in an $SU(3)$ matrix, $\Sigma$. Their masses are incorporated through a scalar-pseudo\-scalar source $\chi=2B_{0}(s+ip)$, where the scalar source, $s$ is equal to the quark mass matrix, $M=\textrm{diag}(m_{u},m_{d},m_{s})$ and the pseudoscalar source $p$ is zero. The external magnetic field is incorporated through left and right sources, which are defined as
\begin{align}
r_{\mu}=l_{\mu}=-eQA_{\mu}^{\rm ext}=-\tfrac{e}{2}\left(\lambda_{3}+\tfrac{1}{\sqrt{3}}\lambda_{8}\right)\ ,
\end{align}
where $Q={\rm diag}(+\tfrac{2}{3},-\tfrac{1}{3},-\tfrac{1}{3})$ is the quark charge matrix, which has been written in terms of the Gell-Mann matrices $\lambda_{3}$ and $\lambda_{8}$ and the corresponding field-strength tensors associated with the left-and-right sources are 
\begin{align}
F_{\mu\nu}^{R}&=\partial_{\mu}r_{\nu}-\partial_{\nu}r_{\mu}-i[r_{\mu},r_{\nu}]&F_{\mu\nu}^{L}=\partial_{\mu}l_{\nu}-\partial_{\nu}l_{\mu}-i[l_{\mu},l_{\nu}]\ ,
\end{align} 
and $A_{\mu}^{\rm ext}$ is the electromagnetic gauge field, which is required to define the covariant derivative that enters the chiral Lagrangian
\begin{align}
\nabla_{\mu}\Sigma=\partial_{\mu}\Sigma-ir_{\mu}\Sigma+i\Sigma l_{\mu}=\partial_{\mu}\Sigma-ieA_{\mu}^{\rm ext}[Q,\Sigma]\ .
\end{align}
The Lagrangian is organized in a power counting scheme where the $SU(3)$ field $\Sigma$ is $\mathcal{O}(p^{0})$, derivatives and covariant derivatives are $\mathcal{O}(p^{1})$ with the external left-and-right fields, $r_{\mu}$ and $l_{\mu}$, and the gauge field counting as $\mathcal{O}(p^{1})$ and the scalar-pseudoscalar source count as $\mathcal{O}(p^{2})$. The field strength tensors are consequently $\mathcal{O}(p^{2})$. We require the $\chi$PT Lagrangian upto $\mathcal{O}(p^{6})$ -- we will use the notation $\mathcal{L}_{n}$ for the Lagrangian at $\mathcal{O}(p^{n})$ in the chiral expansion. The $\mathcal{O}(p^2)$ \textit{Minkowski} space Lagrangian in $\chi$PT is
\begin{align}
\mathcal{L}_{2}&=-\frac{1}{4}F_{\mu\nu}F^{\mu\nu}+\frac{f^{2}}{4}\Tr [\nabla_{\mu}\Sigma(\nabla^{\mu}\Sigma)^{\dagger}  ]+\frac{f^{2}}{4}\Tr[\chi\Sigma^{\dagger}+\Sigma\chi^{\dagger}  ]\ ,
\end{align} 
where $f$ is the bare pion decay constant and $F_{\mu\nu}$ the electromagnetic tensor associated with the external gauge field. 
The $\mathcal{O}(p^{4})$ Lagrangian in \textit{Minkowski} space required for our calculation is 
\begin{equation}
\begin{split}
\mathcal{L}_{4}&=L_{4}\Tr\left [\nabla_{\mu}\Sigma(\nabla^{\mu}\Sigma)^{\dagger}\right ]\Tr(\chi\Sigma^{\dagger}+\chi^{\dagger}\Sigma)+L_{5}\Tr\left [\nabla_{\mu}\Sigma(\nabla^{\mu}\Sigma)^{\dagger}(\chi\Sigma^{\dagger}+\chi^{\dagger}\Sigma) \right ]\\
&+L_{6}\left [\Tr(\chi\Sigma^{\dagger}+\chi^{\dagger}\Sigma)\right ]^{2}+L_{7}\left [\Tr(\chi\Sigma^{\dagger}-\chi^{\dagger}\Sigma)\right ]^{2}+L_{8}\Tr \left(\Sigma \chi^{\dagger}\Sigma \chi^{\dagger}+\chi\Sigma^{\dagger}\chi\Sigma^{\dagger} \right )\\
&-iL_{9}\Tr\left [F^{R}_{\mu\nu}\nabla^{\mu}\Sigma(\nabla^{\nu}\Sigma)^{\dagger}+F^{L}_{\mu\nu}(\nabla^{\mu}\Sigma)^{\dagger}\nabla^{\nu}\Sigma \right ]+L_{10}\Tr\left [\Sigma F_{\mu\nu}^{L}\Sigma^{\dagger}F^{R\mu\nu}\right ]\\
&+H_{1}\Tr\left [F_{\mu\nu}^{R}F^{R\mu\nu}+F_{\mu\nu}^{L}F^{L\mu\nu} \right ]+H_{2}\Tr(\chi \chi^{\dagger})\ ,
\end{split}
\end{equation}
where $L_{i}$ ($H_{i}$) are the low-energy (high-energy) constants, which are defined as
\begin{align}
\label{eq:LiHi}
L_{i}&=L_{i}^{r}+\Gamma_{i}\lambda\ , &
H_{i}&=H_{i}^{r}+\Delta_{i}\lambda\ ,&
\lambda&=-\frac{\Lambda^{-2\epsilon}}{2(4\pi)^{2}}\left(\frac{1}{\epsilon}+1\right)\ ,
\end{align}
where $\Gamma_{i}$ and $\Delta_{i}$ are constants required for renormalization, see Eq.~\eqref{eq:GiDi}. 

The running of the renormalized couplings, $L_{i}^{r}$ and $H_{i}^{r}$, is deduced straightforwardly from their definitions since the bare low-energy and high-energy constant are scale independent,
\begin{align}
\label{eq:Lri}
\Lambda\frac{dL_{i}^{r}}{d\Lambda}&=-\frac{\Gamma_{i}}{(4\pi)^{2}}\ ,&
\Lambda\frac{dH_{i}^{r}}{d\Lambda}&=-\frac{\Delta_{i}}{(4\pi)^{2}}\ ,
\end{align}
and will be important in verifying the scale-invariance of the free energy (density). There are further low-energy constants that enter through the $\mathcal{O}(p^{6})$ $\chi$PT Lagrangian,
\begin{equation}
\begin{split}
\mathcal{L}_{6}&=C_{19}\Tr[(\chi\Sigma^{\dagger}+\chi^{\dagger}\Sigma)^{3}]+C_{20}\Tr[\chi\Sigma^{\dagger}+\chi^{\dagger}\Sigma]^{2}\Tr[\chi\Sigma^{\dagger}+\chi^{\dagger}\Sigma]+C_{21}\Tr[\chi\Sigma^{\dagger}+\chi^{\dagger}\Sigma]^{3}\\
&+C_{61}\Tr[(\Sigma^{\tfrac{1}{2}})^{\dagger}\chi(\Sigma^{\tfrac{1}{2}})^{\dagger}+\Sigma^{\tfrac{1}{2}}\chi^{\dagger}\Sigma^{\tfrac{1}{2}}]+C_{62}\Tr[\chi\Sigma^{\dagger}+\Sigma\chi^{\dagger}]\Tr [f_{+\mu\nu}f_{+}^{\mu\nu}]\\
&+C_{94}\det[\chi+\chi^{\dagger}],
\end{split}
\end{equation}
with only terms relevant to our analysis presented. Magnetic field dependence enters exclusively through $f_{+\mu\nu}$, which is defined in terms of the left-and-right field strength tensors and the $SU(3)$ field, $f_{+\mu\nu}=\Sigma^{\frac{1}{2}}F_{\mu\nu}^{L}(\Sigma^{\frac{1}{2}})^{\dagger}+(\Sigma^{\frac{1}{2}})^{\dagger}F_{\mu\nu}^{R}\Sigma^{\frac{1}{2}}$. The bare coupling constants, $C_{i}$, 
\begin{align}
C_{i}=\frac{D^{r}_{i}}{(4\pi f)^{2}}-\frac{1}{(4\pi)^{2}}\frac{\Gamma_{i}^{(2)}\Lambda^{-4\epsilon}}{4(4\pi f)^{2}}\left(\frac{1}{\epsilon}+1\right)^{2}+\frac{(\Gamma_{i}^{(1)}+\Gamma_{i}^{(L)})\Lambda^{-2\epsilon}}{2(4\pi f)^{2}}\left(\frac{1}{\epsilon}+1\right)
\end{align}
are defined in terms of their renormalized counterparts $D^{r}_{ij}=(4\pi)^{2}C^{r}_{ij}$, the constants $\Gamma_{ij}^{(2)}$ and $\Gamma_{ij}^{(1)}$, see Eqs.~\eqref{eq:G2i}, \eqref{eq:G1i} and \eqref{eq:G1imore}, and $\Gamma_{ij}^{(L)}$ are linear combinations of the renormalized low energy constants, $L^{r}_{i}$ -- we list these in Eqs.~\eqref{eq:GammaL19}-\eqref{eq:GammaL94}. The running of the renormalized couplings $D^{r}_{i}$ are most conveniently expressed in combinations that appear in the $\mathcal{O}(p^{6})$ free energy (density) and the quark condensates, see Eqs.\eqref{eq:Dri-eq1}-\eqref{eq:Dri-eq8}.

The contribution to the free energy at $\mathcal{O}(p^{6})$ arises through two-loop diagrams with vertices from $\mathcal{L}_{2}$, one-loop diagrams with vertices from $\mathcal{L}_{4}$ and tree-level diagrams with vertices from $\mathcal{L}_{6}$. There we need to expand $\mathcal{L}_{2}$ up to quartic order in the meson fields, $\mathcal{L}_{4}$ up to quadratic order and require the tree-level $\mathcal{L}_{6}$ contributions. In order to do so, the $SU(3)$ field is most conveniently parametrized in an exponential representation $\Sigma=\exp\left(\frac{i\phi_{a}\lambda_{a}}{f}\right)$, involving the Gell-Mann matrices, $\lambda_{a}$, and Einstein summation convention for repeated indices is assumed. In an external background magnetic field, it is convenient to work in the basis of the charged eigenstates
\begin{align}
\phi_{a}\lambda_{a}&=
\begin{pmatrix}
\pi^{0}+\tfrac{1}{\sqrt{3}}\eta&\sqrt{2}\pi^{+}&\sqrt{2}K^{+}\\
\sqrt{2}\pi^{-}&-\pi^{0}+\tfrac{1}{\sqrt{3}}\eta&\sqrt{2}K^{0}\\
\sqrt{2}K^{-}&\sqrt{2}\bar{K}^{0}&-\tfrac{2}{\sqrt{3}}\eta
\end{pmatrix}\ .
\end{align}
The static Lagrangian in the isospin limit is
\begin{align}
\mathcal{L}_{2,0}&=-\frac{1}{2}H^{2}+\frac{1}{2}f^{2}(\mathring{m}_{\pi}^{2}+2\mathring{m}_{K}^{2}) \,,
\end{align}
where $H$ is the external magnetic field and the Lagrangian quadratic in the meson fields is
\begin{equation}
\begin{split}
\mathcal{L}_{2,2}&=D_{\mu}\pi^{+}D^{\mu}\pi^{-}-\mathring{m}_{\pi}^{2}\pi^{+}\pi^{-}+D_{\mu}K^{+}D^{\mu}K^{-}-\mathring{m}_{K}^{2}K^{+}K^{-}\\
&+\partial_{\mu}K^{0}\partial^{\mu}\bar{K}^{0}-\mathring{m}_{K}^{2}K^{0}\bar{K}^{0}+\frac{1}{2}\partial_{\mu}\pi^{0}\partial^{\mu}\pi^{0}-\frac{1}{2}\mathring{m}_{\pi}^{2}(\pi^{0})^{2}+\frac{1}{2}\partial_{\mu}\eta\partial^{\mu}\eta-\frac{1}{2}\mathring{m}_{\eta}^{2}\eta^{2}\ ,
\end{split}
\end{equation}
where the covariant derivatives are defined for the charged scalar fields, $\pi^{\pm}$ and $K^{\pm}$ as $D_{\mu}\mathcal{C}^{\pm}=(\partial_{\mu}\pm ieA_{\mu}^{\rm ext})\mathcal{C}^{\pm}$. In the isospin limit, $m_{u}=m_{d}$, there is no mixing between the neutral pion and the eta. The meson octet bare masses, $\mathring{m}_{i}$, in terms of the degenerate light quark mass, $\hat{m}=\frac{1}{2}(m_{u}+m_{d})$, and the strange quark mass $m_{s}$, are
\begin{align}
\label{eq:mesonmasses}
\mathring{m}_{\pi}^{2}&=2B_{0}\hat{m}&
\mathring{m}_{K}^{2}&=B_{0}(\hat{m}+m_{s})&
\mathring{m}_{\eta}^{2}&=\tfrac{2}{3}B_{0}(\hat{m}+2m_{s})\ .
\end{align}
In order to find the vacuum free energy, we require four mesonic field contributions from $\mathcal{L}_{2}$, tree-level and two-mesonic field contributions from $\mathcal{L}_{4}$ and tree-level contributions from $\mathcal{L}_{6}$. We present these in \ref{app:L}.

\section{The charged propagator}
\label{sec:propagator}
For the calculation of the loop contributions to the free energy, we require expressions for the meson propagators in a background magnetic field. We work in \textit{Euclidean} space and choose the fully asymmetric gauge $A_{\mu}^{\rm ext}=(0,-Hx_{2},0,0)$ that allows for the utilization of the harmonic oscillator propagator~\cite{tiburzi2008hadrons}. For charged scalar fields $\phi$ and $\phi^{\dagger}$ (either $\pi^{\pm}$ or $K^{\pm}$ in three-flavor $\chi$PT) with mass $m_{\phi}$ and charge $\pm e$, the quadratic action in \textit{Euclidean} space is
\begin{align}
S_{\rm quad}=\int d^{4}x\ \phi^{\dagger}(x)\left[-D_{\mu}D_{\mu}+m_{\phi}^{2}\right]\phi(x)\ ,
\end{align}
which can be simplified using the Fourier representation of the charged scalar field
\begin{align}
\phi(x)=\int\frac{d^{3}\tilde{k}}{(2\pi)^{3}}e^{- i \tilde{k}\cdot\tilde{x}}\phi(\tilde{k},x_{2})\ ,
\end{align}
where $\tilde{k}=(k_{0},0,k_{2},k_{3})$ and $\tilde{x}=(x_{0},0,x_{2},x_{3})$. Then the quadratic action, after an integration over $\tilde{x}$, which gives rise to $(2\pi)^{3}\delta^{(3)}(\tilde{k}-\tilde{k}')$, a further integration over $\tilde{k}'$, becomes
\begin{align}
S_{\rm quad}&=\int dx_{2}\int \frac{d^{3}\tilde{k}}{(2\pi)^{3}}\ \phi^{\dagger}(\tilde{k},x_{2})\left[2\left(\frac{1}{2}p_{X}^{2}+\frac{1}{2}(eH)^{2}X^{2}+\frac{1}{2}\left[k_{0}^{2}+k_{3}^{2}+m_{\phi}^{2}\right]\right)\right]\phi(\tilde{k},x_{2})\ ,
\end{align}
from which the inverse propagator is easily identified
\begin{align}
\mathcal{D}^{-1}=2\left(\frac{1}{2}p_{X}^{2}+\frac{1}{2}(eH)^{2}X^{2}\right)+\left[k_{0}^{2}+k_{3}^{2}+m_{\phi}^{2}\right]\equiv 2\mathcal{H}+E_{\perp}^{2}\ ,
\end{align}
where $X=x_{2}+\frac{k_{1}}{eH}$, $p_{X}=-i\partial_{X}$ and $E_{\perp}^{2}=k_{0}^{2}+k_{3}^{2}+m_{\phi}^{2}$. The propagator can be recast in the Schwinger proper time form, which in the $|X\rangle$ basis is
\begin{align}
\mathcal{D}(X',X)=\langle X'|\mathcal{D}|X\rangle=\frac{1}{2}\int_{0}^{\infty}ds\ \langle X'|e^{-s\mathcal{H}}|X\rangle e^{-sE_{\perp}^{2}/2}\ .
\end{align}
Utilizing the harmonic oscillator propagator, 
\begin{align}
\langle X'|e^{-s\mathcal{H}}|X\rangle=\sqrt{\frac{eH}{2\pi \sinh eHs}}\exp\left[-\frac{eH}{2\sinh eHs}\left\{(X'^{2}+X^{2})\cosh eHs-2X'X\right\}\right]
\end{align}
and performing the momentum integrals in the quadratic action, which are all Gaussian, the propagator in position space (after a change of variables $s\rightarrow 2s$) is 
\begin{align}
\mathcal{D}(x',x)=e^{ieH\Delta x_{1}\bar{x}_{2}}\frac{1}{(4\pi)^{2}}\int_{0}^{\infty}\frac{ds}{s^{2}}\frac{eHs}{\sinh eHs}e^{-m_{\phi}^{2}s}\exp\left[-\tfrac{eH\left(\Delta x_{1}^{2}+\Delta x_{2}^{2}\right)}{4\tanh eHs}-\tfrac{\Delta x_{0}^{2}+\Delta x_{3}^{2}}{4s}\right]\ ,
\end{align}
with $\Delta x_{\mu}=x'_{\mu}-x_{\mu}$ characterizing the difference between  Euclidean coordinates. The propagator satisfies the Green's function identity
\begin{align}
(-D'_{\mu}D'_{\mu}+m_{\phi}^{2})\mathcal{D}(x',x)=\delta^{(4)}(x'-x)\ ,
\end{align}
as is seen by first writing $m_{\phi}^{2}\mathcal{D}(x',x)$ in terms of a proper-time derivative of $e^{-m_{\phi}^{2}s}$ in the integrand. An integration by parts then produces a boundary term with a non-vanishing contribution in the $s\rightarrow 0$ limit, which is a Gaussian representation of a four-dimensional $\delta$-function in Euclidean space. The non-boundary term cancels exactly with the term that arises through the double-covariant derivative in the first variable of $\mathcal{D}(x',x)$.

For coincident points, the propagator is coordinate-independent and can be separated into a divergent $H=0$ contribution and a finite $H$ contribution
\begin{align}
\label{eq:D}
\mathcal{D}(x,x)\equiv \mathcal{D}(m_{\phi}^{2})=\mathcal{D}_{0}(m_{\phi}^{2})+\mathcal{D}_{H}(m_{\phi}^{2})\ .
\end{align}
The divergence can be handled in dimensional regularization in $4-2\epsilon$ dimension that results in the standard expression
\begin{align}
\mathcal{D}_{0}(m_{\phi}^{2})&=-\frac{m_{\phi}^{2}}{(4\pi)^{2}}\left[\frac{1}{\epsilon}+1+\log\frac{\Lambda^{2}}{m_{\phi}^{2}}\right]\ ,
\end{align}
consisting of a chiral log and a dependence on the $\overline{\rm MS}$ scale $\Lambda$. The finite magnetic field-dependent contribution
\begin{align}
\mathcal{D}_{H}(m_{\phi}^{2})=\frac{eH}{(4\pi)^{2}}\mathcal{I}_{H,2}(\tfrac{m_{\phi}^{2}}{eH})\ 
\end{align}
is best expressed in terms of a dimensionless function $\mathcal{I}_{H,2}(z)$, which has a closed form expression~\cite{gradshteyn2014table} that depends on the $\Gamma$-function,
\begin{align}
\label{eq:IH2GR}
\mathcal{I}_{H,2}(z)=2\log\Gamma(\tfrac{1+z}{2})+z-z\log\tfrac{z}{2}-\log(2\pi)\ .
\end{align}

When evaluating vacuum diagrams that contribute to the free energy, the $\mathcal{O}(p^{4})$ contribution is the standard one-loop effective potential. At $\mathcal{O}(p^{6})$, on the other hand, there are diagrams with one or two loops that consist of operators with either one or two covariant derivatives in each loop,
\begin{align}
\label{eq:pDp}
\lim_{x'\rightarrow x}\langle\phi^{\dagger}(x')D_{\mu}\phi(x)\rangle&=\lim_{x'\rightarrow x}D_{\mu}\mathcal{D}(x',x)\ ,\\
\label{eq:DpDp}
\lim_{x'\rightarrow x}\langle D'_{\mu}\phi^{\dagger}(x')D_{\mu}\phi(x)\rangle&=\lim_{x'\rightarrow x}D'_{\mu}D_{\mu}\mathcal{D}(x',x)\ ,
\end{align}
with the covariant derivative taken prior to the coincident limit. There are corresponding operators for neutral mesons with covariant derivatives replaced by regular ones.  Here, we focus only on the charged fields, since unlike their neutral counterparts, the contributions of the charged fields do not vanish trivially. The  vacuum graph containing Eq.~(\ref{eq:pDp}) is non-vanishing in the first spatial direction and an odd function in $x_{2}$. In a one-loop diagram, the contribution vanishes upon integration but in a two-loop diagram with a second charged meson loop, the overall contribution is even in $x_{2}$ but ultimately cancel to zero -- contributions from $\mathcal{L}_{2,4}$ proportional to $D_{\mu}\phi^{\dagger}D_{\mu}\phi\phi^{\dagger}\phi$ cancel with those proportional to $D_{\mu}\phi^{\dagger}D_{\mu}\phi^{\dagger}\phi\phi+D_{\mu}\phi D_{\mu}\phi\phi^{\dagger}\phi^{\dagger}$. The contribution with a single covariant derivative acting on the primed coordinate vanishes trivially in the coincident limit. For the neutral mesons, both single derivative operators vanish but for the charged mesons, the result is asymmetric due to the presence  of the Schwinger phase factor, which produces a minus sign upon the exchange of the first spatial coordinates.

The latter contribution can be handled within one and two-loop vacuum diagrams through integration by parts, which amounts to utilizing the identity
\begin{align}
\lim_{x'\rightarrow x}\langle D'_{\mu}\phi^{\dagger}(x')D_{\mu}\phi(x)\rangle=-m_{\phi}^{2}\mathcal{D}(x,x)\ .
\end{align}
The boundary term that arises vanishes either because it is spatially independent or because the integral is odd in $x_{2}$. A non-boundary term proportional to the $\delta$-function vanishes in dimensional regularization since the resulting integrand is coordinate-independent.

\section{Vacuum free energy, quark condensate shifts and renormalized magnetization at $T=0$}
\label{sec:thermodynamics}
\subsection{Vacuum Free Energy (Density)}
The vacuum free energy, $\mathcal{F}$, relates to the partition function, $Z$ through a proportionality constant that depends on the physical volume $V$ and the inverse temperature $\be$   
\begin{align}
\mathcal{F}=-\frac{1}{\beta V}\ln Z=\sum_{n}\mathcal{F}^{(n)}=\sum_{n}(\mathcal{F}^{(n)}_{0}+\mathcal{F}^{(n)}_{H})\ .
\end{align}
In the second equality, $\mathcal{F}^{(n)}$ is the contribution at $\mathcal{O}(p^{n})$ that in the third equality has been separated into contributions independent of and dependent on the external magnetic field.
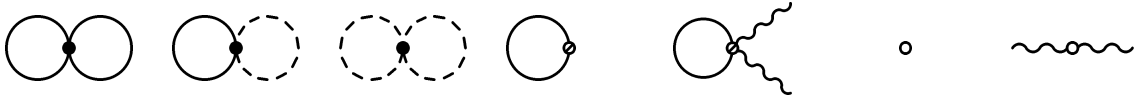
\begin{figure}[t!]
\begin{center}
\unitlength=0.6mm
\begin{fmffile}{vac-cc}
    \begin{fmfgraph*}(30,20)
    	\fmfkeep{vac-cc}
        \fmfleft{i}
        \fmfright{o}
        \fmf{phantom,tension=10}{i,i1}
        \fmf{phantom,tension=10}{o,o1}
        \fmf{plain,left,tension=0.4}{i1,v1,i1}
        \fmf{plain,right,tension=0.4}{o1,v1,o1}
        \fmfdot{v1}
        \fmflabel{$\Sigma$}{v1}
    \end{fmfgraph*}
\end{fmffile}
\begin{fmffile}{vac-cn}
    \begin{fmfgraph*}(30,20)
	\fmfkeep{vac-cn}
        \fmfleft{i}
        \fmfright{o}
        \fmf{phantom,tension=10}{i,i1}
        \fmf{phantom,tension=10}{o,o1}
        \fmf{plain,left,tension=0.4}{i1,v1,i1}
        \fmf{dashes,right,tension=0.4}{o1,v1,o1}
        \fmfdot{v1}
    \end{fmfgraph*}
\end{fmffile}
\begin{fmffile}{vac-nn}
    \begin{fmfgraph*}(30,20)
        \fmfkeep{vac-nn}
        \fmfleft{i}
        \fmfright{o}
        \fmf{phantom,tension=10}{i,i1}
        \fmf{phantom,tension=10}{o,o1}
        \fmf{dashes,left,tension=0.4}{i1,v1,i1}
        \fmf{dashes,right,tension=0.4}{o1,v1,o1}
        \fmfdot{v1}
    \end{fmfgraph*}
\end{fmffile}
\begin{fmffile}{vac-c}
    \begin{fmfgraph*}(30,20)
        \fmfkeep{vac-c}
        \fmfleft{i}
        \fmfright{o}
        \fmf{phantom,tension=10}{i,i1}
        \fmf{phantom,tension=10}{o,o1}
        \fmf{plain,left,tension=0.4}{i1,v1,i1}
        \fmf{phantom,right,tension=0.4}{o1,v1,o1}
        \fmfv{decor.shape=circle,decor.filled=shaded, decor.size=2thick}{v1}
    \end{fmfgraph*}
\end{fmffile}
\begin{fmffile}{vac-cext}
    \begin{fmfgraph*}(30,20)
        \fmfkeep{vac-cext}
        \fmfleft{i}
        \fmfright{oup,odown}
        \fmf{phantom,tension=10}{i,i1}
        \fmf{plain,left,tension=0.4}{i1,v1,i1}
        \fmf{wiggly,tension=0.4}{oup,v1}
        \fmf{wiggly,tension=0.4}{odown,v1}
        \fmfv{decor.shape=circle,decor.filled=shaded, decor.size=2thick}{v1}
    \end{fmfgraph*}
\end{fmffile}
\begin{fmffile}{vac-dot}
    \begin{fmfgraph*}(30,20)
        \fmfkeep{vac-dot}
        \fmfleft{i}
        \fmfright{o}
        \fmf{phantom,tension=10}{i,i1}
        \fmf{phantom,tension=10}{o1,o}
        \fmf{phantom,left,tension=0.4}{i1,v1,i1}
        \fmf{phantom,right,tension=0.4}{o1,v1,o1}
        \fmfv{decor.shape=circle,decor.filled=empty, decor.size=2thick}{v1}
    \end{fmfgraph*}
\end{fmffile}
\begin{fmffile}{vac-tree}
    \begin{fmfgraph*}(30,20)
        \fmfkeep{vac-tree}
        \fmfleft{i}
        \fmfright{o}
        \fmf{phantom,tension=10}{i,i1}
        \fmf{phantom,tension=10}{o,o1}
        \fmf{phantom,left,tension=0.4}{i1,v1,i1}
        \fmf{phantom,right,tension=0.4}{o1,v1,o1}
        \fmf{wiggly,tension=0.4}{i1,v1,o1}
        \fmfv{decor.shape=circle,decor.filled=empty, decor.size=2thick}{v1}
    \end{fmfgraph*}
\end{fmffile}
\end{center}
\caption{Next-to-next-to-leading order graphs that contribute to the vacuum free energy (density). The solid lines represent charged mesons, the dashed lines represent neutral mesons and the wiggly line represents a magnetic field insertion. The solid vertex contributes at $\mathcal{O}(p^{2})$, the dashed vertex at $\mathcal{O}(p^{4})$ and the empty vertex at $\mathcal{O}(p^{6})$.}
\label{fig:vacuumgraphs}
\end{figure}

The contributions to $\mathcal{F}^{(2)}$ arises through $\mathcal{L}_{2,0}$ and consists of a pure gauge contribution, $\mathcal{F}^{(2)}_{H}=\tfrac{1}{2}H^{2}$ and a magnetic field independent contribution, $\mathcal{F}^{(2)}_{0}=\frac{f^{2}}{2}(\mathring{m}_{\pi}^{2}+2\mathring{m}_{K}^{2})$. 
Contributions from $\mathcal{F}^{(4)}$ and $\mathcal{F}^{(6)}$ renormalize the magnetic field. At $\mathcal{O}(p^{4})$, there are contributions that arise through the charged pions, charged kaons, neutral kaons, neutral pion and eta loops, 
\begin{align}
\mathcal{F}^{(4)}_{1}&=I_{H}(\mathring{m}_{\pi})+I_{H}(\mathring{m}_{K})+I_{0}(\mathring{m}_{K})+\frac{1}{2}I_{0}(\mathring{m}_{\pi})+\frac{1}{2}I_{0}(\mathring{m}_{\eta})
\end{align}
respectively. The integral $I_{H}(m_{\phi})$ has a Schwinger proper-time representation
\begin{align}
\label{eq:IH}
I_{H}(m_{\phi})=-\frac{(e^{\gamma_{E}}\Lambda^{2})^\epsilon}{(4\pi)^{2}}\int_{0}^{\infty}\frac{1}{s^{3-\epsilon}}e^{-m_{\phi}^{2}s}\left[\frac{eHs}{\sinh eHs}\right]\ ,
\end{align}
which contains divergences proportional to quintic power in the meson mass and another that is quadratic in the external magnetic field. These are cancelled by the tree-level counter-term contribution, $\mathcal{F}^{(4)}_{\rm ct}=-\mathcal{L}_{4,0}$. 

The vacuum diagrams that contribute to $\mathcal{F}^{(6)}$ arise through two-loop diagrams of $\mathcal{L}_{2,4}$, one-loop diagrams of $\mathcal{L}_{4,2}$ and tree diagrams of $\mathcal{L}_{6,0}$ as depicted in Fig.~\ref{fig:vacuumgraphs}. The two-loop diagrams containing two charged pion loops, two charged kaon loops, and a charged pion loop and a charged kaon loop, each vanish separately,
\begin{align}
\mathcal{F}_{2}^{(6)}[\ \parbox{20mm}{\fmfreuse{vac-cc}}]=0\ .
\end{align}
This is deduced from the third and fourth lines in $\mathcal{L}_{2,4}$ -- upon utilizing the identity of Eq.~(\ref{eq:DpDp}), one notes that the coefficients of each of the contributions proportional to $\mathcal{D}(\mathring{m}_{i})\mathcal{D}(\mathring{m}_{j})$, where the masses are either that of the charged pions or the charged kaons. The cancellation appears accidental and also occurs in two-flavor $\chi$PT. The contribution of two-loop vacuum diagrams containing at least one neutral meson loop is non-vanishing. These are deduced similarly using the remaining terms in $\mathcal{L}_{2,4}$,
\begin{align}
\mathcal{F}_{2}^{(6)}[\ \parbox{20mm}{\fmfreuse{vac-cn}}]&=\frac{\mathring{m}_{\pi}^{2}}{f^{2}}\left[\frac{1}{2}\mathcal{D}(\mathring{m}_{\pi})\mathcal{D}_{0}(\mathring{m}_{\pi})-\frac{1}{6}\mathcal{D}(\mathring{m}_{\pi})\mathcal{D}_{0}(\mathring{m}_{\eta})\right]+\frac{\mathring{m}_{K}^{2}}{f^{2}}\left[\frac{1}{3}\mathcal{D}(\mathring{m}_{K})\mathcal{D}_{0}(\mathring{m}_{\eta})\right]\\ \nonumber
\mathcal{F}_{2}^{(6)}[\ \parbox{20mm}{\fmfreuse{vac-nn}}]&=\frac{\mathring{m}_{\pi}^{2}}{f^{2}}\left[-\frac{1}{8}\mathcal{D}_{0}(\mathring{m}_{\pi})^{2}-\frac{1}{12}\mathcal{D}_{0}(\mathring{m}_{\pi})\mathcal{D}_{0}(\mathring{m}_{\eta})+\frac{7}{12}\mathcal{D}_{0}(\mathring{m}_{\eta})^{2}\right]\\
&+\frac{\mathring{m}_{K}^{2}}{f^{2}}\left[\frac{1}{3}\mathcal{D}_{0}(\mathring{m}_{K})\mathcal{D}_{0}(\mathring{m}_{\eta})-\frac{2}{9}\mathcal{D}_{0}(\mathring{m}_{\eta})^{2}\right]\ .
\end{align}
and so are the single meson loop contributions that arise through $\mathcal{L}_{4,2}$ and terms proportional to the low energy constants $L_{4}$ through $L_{8}$,
\begin{align}
\nonumber
&\mathcal{F}^{(6)}[\ \parbox{20mm}{\fmfreuse{vac-c}}]\\ \nonumber
=&\frac{4L_{4}}{f^{2}}(\mathring{m}_{\pi}^{2}+2\mathring{m}_{K}^{2})[\mathring{m}_{\pi}^{2}\{2\mathcal{D}(\mathring{m}_{\pi})+\mathcal{D}_{0}(\mathring{m}_{\pi})\}+2\mathring{m}_{K}^{2}\{\mathcal{D}(\mathring{m}_{K})+\mathcal{D}_{0}(\mathring{m}_{K})\}+\mathring{m}_{\eta}^{2}\mathcal{D}_{0}(\mathring{m}_{\eta})]\\ \nonumber
+&\frac{4L_{5}}{f^{2}}[\mathring{m}_{\pi}^{4}\{2\mathcal{D}(\mathring{m}_{\pi})+\mathcal{D}_{0}(\mathring{m}_{\pi})\}+2\mathring{m}_{K}^{4}\{\mathcal{D}(\mathring{m}_{K})+\mathcal{D}_{0}(\mathring{m}_{K})\}+\mathring{m}_{\eta}^{2}\mathcal{D}_{0}(\mathring{m}_{\eta})]\\ \nonumber
+&\frac{8L_{6}}{f^{2}}(\mathring{m}_{\pi}^{2}+2\mathring{m}_{K}^{2})[\mathring{m}_{\pi}^{2}\{2\mathcal{D}(\mathring{m}_{\pi})+\mathcal{D}_{0}(\mathring{m}_{\pi})\}+2\mathring{m}_{K}^{2}\{\mathcal{D}(\mathring{m}_{K})+\mathcal{D}_{0}(\mathring{m}_{K})\}+\mathring{m}_{\eta}^{2}\mathcal{D}_{0}(\mathring{m}_{\eta})]\\ \nonumber
+&\frac{64L_{7}}{3f^{2}}(\mathring{m}_{\pi}^{2}-\mathring{m}_{K}^{2})^{2}\mathcal{D}_{0}(\mathring{m}_{\eta})+\frac{16L_{8}}{f^{2}}[\mathring{m}_{\pi}^{4}\mathcal{D}(\mathring{m}_{\pi})+\tfrac{1}{2}\mathring{m}_{\pi}^{4}\mathcal{D}_{0}(\mathring{m}_{\pi})+\mathring{m}_{K}^{4}\{\mathcal{D}(\mathring{m}_{K})+\mathcal{D}_{0}(\mathring{m}_{\pi})\}\\
+&\tfrac{1}{3}(4\mathring{m}_{K}^{4}-4\mathring{m}_{\pi}^{2}\mathring{m}_{K}^{2}+\tfrac{3}{2}\mathring{m}_{\pi}^{4})\mathcal{D}_{0}(\mathring{m}_{\eta})]\ .
\end{align}
The terms proportional to $L_{9}$ and $L_{10}$ contain two external magnetic field insertions each. For the former, this is straightforward to note from the Lagrangian but for the latter an integration by parts is required. Assuming a uniform magnetic field the contribution is proportional to $[D_{x},D_{y}]=-ieH$ and the contribution to the free energy contains two external magnetic field insertions and a charged propagator each for the pion and the kaon,
\begin{align}
\mathcal{F}^{(6)}[\ \parbox{20mm}{\fmfreuse{vac-cext}}]&=\frac{4(eH)^{2}}{f^{2}}(L_{9}+L_{10})[\mathcal{D}(\mathring{m}_{\pi})+\mathcal{D}(\mathring{m}_{K})]\ .
\end{align}
The final $\mathcal{O}(p^{6})$ contributions are tree-level counter-terms equal to negative of $\mathcal{L}_{6,0}$ and consists of two types of contributions, one of which is absent of external field insertions,
\begin{align}
\nonumber
\mathcal{F}^{(6)}[\ \parbox{20mm}{\fmfreuse{vac-dot}}]=&-2\mathring{m}_{\pi}^{6}[4C_{19}+12C_{20}+4C_{21}-C_{94}]\\ \nonumber
&-4\mathring{m}_{\pi}^{4}\mathring{m}_{K}^{2}[12C_{19}+4C_{20}+12C_{21}+C_{94}]\\ 
&+32\mathring{m}_{\pi}^{2}\mathring{m}_{K}^{4}[3C_{19}+C_{20}-3C_{21}]-64\mathring{m}_{K}^{6}[C_{19}+C_{20}+C_{21}]\\
\mathcal{F}^{(6)}[\ \parbox{20mm}{\fmfreuse{vac-tree}}]=&-\frac{32}{9}(eH)^{2}\mathring{m}_{\pi}^{2}[2C_{61}+3C_{62}]-\frac{32}{9}(eH)^{2}\mathring{m}_{K}^{2}[C_{61}+6C_{62}]\ .
\end{align}
The single poles in the external field dependent contribution cancel with the single poles in the contribution proportional to $(eH)^{2}(L_{9}+L_{10})$. The remaining divergences are single and double poles arising out of the charged pion-neutral meson double bubble and the charged kaon-neutral meson double bubble, which cancel with the respective one-loop diagrams (arising through $\mathcal{L}_{4,2}$) involving charged pions or kaons. The full renormalized $\mathcal{O}(p^{6})$ vacuum free energy (density), $\mathcal{F}$, consists of a magnetic field independent contribution, $\mathcal{F}_{0}$, which is presented in \ref{app:freeenergy}, while the field-dependent contribution, $\mathcal{F}_{H}$, is

\begin{align}
\nonumber
\mathcal{F}_{H}&=\frac{1}{2}H_{R}^{2}+\frac{(eH)^{2}}{(4\pi)^{2}}\left[\mathfrak{I}_{H}(\tfrac{\mathring{m}^{2}_{\pi}}{eH})+\mathfrak{I}_{H}(\tfrac{\mathring{m}^{2}_{K}}{eH})\right]\\
\nonumber
&-\frac{\mathring{m}_{\pi}^{4}(eH)}{(4\pi f)^{2}}\left[\frac{1}{2(4\pi)^{2}}\log\frac{\Lambda^{2}}{\mathring{m}_{\pi}^{2}}+\frac{1}{18(4\pi)^{2}}\log\frac{\Lambda^{2}}{\mathring{m}_{\eta}^{2}}+8(L^{r}_{4}+L^{r}_{5})-16(L^{r}_{6}+L^{r}_{8})\right]\mathcal{I}_{H,2}(\tfrac{\mathring{m}^{2}_{\pi}}{eH})\\
\nonumber
&-\frac{\mathring{m}_{\pi}^{2}\mathring{m}_{K}^{2}(eH)}{(4\pi f)^{2}}\left[-\frac{1}{9(4\pi)^{2}}\log\frac{\Lambda^{2}}{\mathring{m}_{\eta}^{2}}+8(L^{r}_{4}-2L^{r}_{6})\right]\left[2\mathcal{I}_{H,2}(\tfrac{\mathring{m}^{2}_{\pi}}{eH})+\mathcal{I}_{H,2}(\tfrac{\mathring{m}^{2}_{K}}{eH})\right]\\
\nonumber
&-\frac{\mathring{m}_{K}^{4}(eH)}{(4\pi f)^{2}}\left[\frac{4}{9(4\pi)^{2}}\log\frac{\Lambda^{2}}{\mathring{m}_{\eta}^{2}}+8(2L^{r}_{4}+L^{r}_{5})-16(2L^{r}_{6}+L^{r}_{8})\right]\mathcal{I}_{H,2}(\tfrac{\mathring{m}^{2}_{K}}{eH})\\
\label{eq:FH}
&+\frac{4(eH)^{3}}{(4\pi f)^{2}}(L^{r}_{9}+L^{r}_{10})\left[\mathcal{I}_{H,2}(\tfrac{\mathring{m}^{2}_{\pi}}{eH})+\mathcal{I}_{H,2}(\tfrac{\mathring{m}^{2}_{K}}{eH})\right]\ .
\end{align}
Here, $H_{R}=Z_{H}H$ is the renormalized magnetic field with
\begin{align}
\nonumber
Z_{H}=&\left[1-\frac{4e^{2}}{3}(L^{r}_{10}+2H^{r}_{1})+\frac{e^{2}}{6(4\pi)^{2}}\left(\log\frac{\Lambda^{2}}{\mathring{m}_{\pi}^{2}}+\log\frac{\Lambda^{2}}{\mathring{m}_{K}^{2}}-2\right)\right.\\ \nonumber
&\left.-\frac{e^{2}\mathring{m}_{\pi}^{2}}{2(4\pi f)^{2}}\left\{\frac{32}{9}(2D^{r}_{61}+3D^{r}_{62})+4(L^{r}_{9}+L^{r}_{10})\log\frac{\Lambda^{2}}{\mathring{m}_{\pi}^{2}}\right\}\right.\\
\label{eq:ZH}
&\left.-\frac{e^{2}\mathring{m}_{K}^{2}}{2(4\pi f)^{2}}\left\{\frac{32}{9}(D^{r}_{61}+6D^{r}_{62})+4(L^{r}_{9}+L^{r}_{10})\log\frac{\Lambda^{2}}{\mathring{m}_{K}^{2}}\right\}\right]\ .
\end{align}
A corresponding renormalization of the pion and kaon charge, i.e. $e_{R}=Z_{H}^{-1}e$ ensures the product $eH$, through which the magnetic field contribution enters the matter contribution to the free energy is scale independent. This finite renormalization procedure was first adopted by Schwinger~\cite{Schwinger:1951nm} and deserves further discussion~\cite{endrHodi2013qcd,Adhikari:2021bou}. All contributions that are quadratic in the external field have been absorbed into the pure gauge contribution of the free energy. This choice is not unique since one is free to add and subtract finite pieces \textit{ad hoc} under the constraint that the total free energy remains unaffected. Preference for Schwinger's scheme is based on its physical virtue: it separates the contribution to the free energy into a pure gauge that incorporates all the quadratic contributions, while the pure hadronic contribution incorporates the free energy associated with virtual hadron loops interacting with the external field. This creates current loops that magnetize the (quantum) field theoretic vacuum~\cite{Adhikari:2021bou} and on physical grounds are expected to vanish in the infinite mass limit. On the other hand, this hadronic contribution to the magnetization, later referred to as the \textit{renormalized magnetization}, asymptotes to infinity as the mass of any pair of the charged mesons approaches zero, again a physically virtuous result.

Prior to discussing the effect of the magnetic field on the condensates and the free energy (through the renormalized magnetization), it is worth noting that the free energy is a scale-indepen\-dent quantity, as can be verified by utilizing the running of the renormalized low and high-energy constants of Eq.~\eqref{eq:Lri} and Eqs.~\eqref{eq:Dri-eq1}-\eqref{eq:Dri-eq6}. We have arranged the various contributions to the free energy such that each line in Eq.~\eqref{eq:FH} is independently scale-invariant. The same is true for $Z_{H}$ in Eq.~\eqref{eq:ZH}. Its scale-invariance ensures that both the renormalized electric charge and renormalized magnetic field are separately scale-invariant.
\subsection{Quark Condensate Shifts}
The mass term of the QCD Lagrangian, in the isospin limit, $\mathcal{L}_{\rm mass}=-[\hat{m}(\bar{u}u+\bar{d}d)+m_{s}\bar{s}s]$ , permits the calculation of the total light quark condensate, $\langle \bar{q}q\rangle$, the sum of the up-and-down quark condensates, which are degenerate in the isospin limit, and the the strange quark condensate, $\langle \bar{s}s\rangle$. The former is defined as the differential change in the free energy when the average light quark mass, $\hat{m}$ is altered while the latter is analogously defined for differential changes in the strange quark mass, $m_{s}$,
\begin{align}
\label{eq:qqss}
\langle \bar{q}q\rangle&=\frac{\partial\mathcal{F}}{\partial\hat{m}}\ ,&\langle \bar{s}s\rangle&=\frac{\partial\mathcal{F}}{\partial m_{s}}\ .
\end{align}
Here we focus on the study of the shift induced by the external magnetic field using PDG parameters and also compare our results to those from a recent lattice study~\cite{ding2020chiral}. Since the effect of the external field enters through the interaction of virtual (charged) mesons with the magnetic background, the condensate shift first appears at $\mathcal{O}(p^{4})$. Then,
\begin{align}
\langle\bar{q}q\rangle_{H}&=\langle\bar{q}q\rangle_{H}^{(4)}+\langle\bar{q}q\rangle_{H}^{(6)}\ ,&\langle\bar{s}s\rangle_{H}&=\langle\bar{s}s\rangle_{H}^{(4)}+\langle\bar{s}s\rangle_{H}^{(6)}\ ,
\end{align}
where the $\mathcal{O}(p^{4})$ shift of the condensates is
\begin{align}
\langle\bar{q}q\rangle_{H}^{(4)}&=-\frac{B_{0}(eH)}{(4\pi)^{2}}[2\mathcal{I}_{H,2}(\tfrac{\mathring{m}_{\pi}^{2}}{eH})+\mathcal{I}_{H,2}(\tfrac{\mathring{m}_{K}^{2}}{eH})]\ ,&
\langle\bar{s}s\rangle_{H}^{(4)}&=-\frac{B_{0}(eH)}{(4\pi)^{2}}[\mathcal{I}_{H,2}(\tfrac{\mathring{m}_{K}^{2}}{eH})]\ .
\end{align}
The light quark condensate depends on the charged pion and kaon masses with the factor of two explained by the number of valence up-and-down quarks or anti-quarks in the charged pions compared to the charged kaons. The contribution of the charged kaon to the condensates is identical. The $\mathcal{O}(p^{6})$ contribution to the condensates are rather lengthy compared to two-flavor calculations. Nevertheless, we present them below for completeness,
\begin{align} \nonumber
\langle\bar{q}q\rangle_{H}^{(6)}=&-\frac{B_{0}(eH)^{2}}{(4\pi f)^{2}}\Bigg[\frac{160}{9}D^{r}_{61}+\frac{128}{3}D^{r}_{62}\Bigg.\\
\nonumber
&\left.-4(L^{r}_{9}+L^{r}_{10})\left(3-2\log\frac{\Lambda^{2}}{\mathring{m}_{\pi}^{2}}-\log\frac{\Lambda^{2}}{\mathring{m}_{K}^{2}}-2(eH)\{\mathcal{I}_{H,1}(\tfrac{\mathring{m}_{\pi}^{2}}{eH})+\mathcal{I}_{H,1}(\tfrac{\mathring{m}_{K}^{2}}{eH})\}\right)\right]\\ \nonumber
&+\frac{B_{0}\mathring{m}_{\pi}^{2}(eH)}{(4\pi)^{2}f^{2}}\left[\frac{8}{9(4\pi)^{2}}
-\frac{2}{(4\pi)^{2}}\log\frac{\Lambda^{2}}{\mathring{m}_{\pi}^{2}}-48L^{r}_{4}
-32L^{r}_{5}+96L^{r}_{6}+64L^{r}_{8}\right]\mathcal{I}_{H,2}(\tfrac{\mathring{m}_{\pi}^{2}}{eH})\\ \nonumber
&+\frac{B_{0}\mathring{m}_{\pi}^{2}(eH)}{(4\pi f)^{2}}\left[
\frac{1}{9(4\pi)^{2}}\log\frac{\Lambda^{2}}{\mathring{m}_{\eta}^{2}}-8L^{r}_{4}+16L^{r}_{6}\right]\mathcal{I}_{H,2}(\tfrac{\mathring{m}_{K}^{2}}{eH})\\ \nonumber
&+\frac{B_{0}\mathring{m}_{K}^{2}(eH)}{(4\pi f)^{2}}\left[\frac{4}{9(4\pi)^{2}}\log\frac{\Lambda^{2}}{\mathring{m}_{\eta}^{2}}-32L^{r}_{4}+64L^{r}_{6}\right]\mathcal{I}_{H,2}(\tfrac{\mathring{m}_{\pi}^{2}}{eH})\\ \nonumber
&+\frac{B_{0}\mathring{m}_{K}^{2}(eH)}{(4\pi f)^{2}}\left[\frac{2}{9(4\pi)^{2}}-\frac{2}{3(4\pi)^{2}}\log\frac{\Lambda^{2}}{\mathring{m}_{\eta}^{2}}-48L^{r}_{4}-16L^{r}_{5}+96L^{r}_{6}+32L^{r}_{8}\right]\mathcal{I}_{H,2}(\tfrac{\mathring{m}_{K}^{2}}{eH})\\ \nonumber
&+\frac{B_{0}\mathring{m}_{\pi}^{4}(eH)}{(4\pi f)^{2}}\left[\frac{1}{(4\pi)^{2}}\log\frac{\Lambda^{2}}{\mathring{m}_{\pi}^{2}}+\frac{1}{9(4\pi)^{2}}\log\frac{\Lambda^{2}}{\mathring{m}_{\eta}^{2}}+16L^{r}_{4}+16L^{r}_{5}-32L^{r}_{6}-32L^{r}_{8}\right]\mathcal{I}_{H,1}(\tfrac{\mathring{m}_{\pi}^{2}}{eH})\\ \nonumber
&+\frac{B_{0}\mathring{m}_{\pi}^{2}\mathring{m}_{K}^{2}(eH)}{(4\pi f)^{2}}\left[
-\frac{1}{9(4\pi)^{2}}\log\frac{\Lambda^{2}}{\mathring{m}_{\eta}^{2}}+8L^{r}_{4}-16L^{r}_{6}\right][4\mathcal{I}_{H,1}(\tfrac{\mathring{m}_{\pi}^{2}}{eH})+\mathcal{I}_{H,1}(\tfrac{\mathring{m}_{K}^{2}}{eH})]\\
&+\frac{B_{0}\mathring{m}_{K}^{4}(eH)}{(4\pi f)^{2}}\left[
\frac{4}{9(4\pi)^{2}}\log\frac{\Lambda^{2}}{\mathring{m}_{\eta}^{2}}+16L^{r}_{4}+8L^{r}_{5}
-32L^{r}_{6}-16L^{r}_{8}\right]\mathcal{I}_{H,1}(\tfrac{\mathring{m}_{K}^{2}}{eH})\ ,
\label{eq:qq6}
\end{align}
and
\begin{align}\nonumber
\langle\bar{s}s\rangle_{H}^{(6)}=&-\frac{B_{0}(eH)^{2}}{(4\pi f)^{2}}\left[
\frac{32}{9}D^{r}_{61}+\frac{64}{3}D^{r}_{62}-4(L^{r}_{9}+L^{r}_{10})\left\{1-
\log\frac{\Lambda^{2}}{\mathring{m}_{K}^{2}}-(eH)\mathcal{I}_{H,1}(\tfrac{\mathring{m}_{K}^{2}}{eH})\right\}\right]\\ \nonumber
&+\frac{B_{0}\mathring{m}_{\pi}^{2}(eH)}{(4\pi f)^{2}}\left[-\frac{2}{9(4\pi)^{2}}+
\frac{2}{9(4\pi)^{2}}\log\frac{\Lambda^{2}}{\mathring{m}_{\eta}^{2}}-16L^{r}_{4}+32L^{r}_{6}\right]\mathcal{I}_{H,2}(\tfrac{\mathring{m}_{\pi}^{2}}{eH})\\ \nonumber
&+\frac{B_{0}\mathring{m}_{\pi}^{2}(eH)}{(4\pi f)^{2}}\left[\frac{1}{9(4\pi)^{2}}
\log\frac{\Lambda^{2}}{\mathring{m}_{\eta}^{2}}-8L^{r}_{4}+16L^{r}_{6}\right]
\mathcal{I}_{H,2}(\tfrac{\mathring{m}_{K}^{2}}{eH})\\ \nonumber
&+\frac{B_{0}\mathring{m}_{K}^{2}(eH)}{(4\pi f)^{2}}\left[\frac{4}{9(4\pi)^{2}}-\frac{8}{9(4\pi)^{2}}\log\frac{\Lambda^{2}}{\mathring{m}_{\eta}^{2}}-32L^{r}_{4}-16L^{r}_{5}+64L^{r}_{6}+32L^{r}_{8}\right]\mathcal{I}_{H,2}(\tfrac{\mathring{m}_{K}^{2}}{eH})\\ \nonumber
&+\frac{B_{0}\mathring{m}_{\pi}^{2}\mathring{m}_{K}^{2}(eH)}{(4\pi f)^{2}}\left[-\frac{1}{9(4\pi)^{2}}\log\frac{\Lambda^{2}}{\mathring{m}_{\eta}^{2}}+8L^{r}_{4}-16L^{r}_{6}\right]\mathcal{I}_{H,1}(\tfrac{\mathring{m}_{K}^{2}}{eH})\\ 
&+\frac{B_{0}\mathring{m}_{K}^{4}(eH)}{(4\pi f)^{2}}\left[\frac{4}{9(4\pi)^{2}}\log\frac{\Lambda^{2}}{\mathring{m}_{\eta}^{2}}+16L^{r}_{4}+8L^{r}_{5}-32L^{r}_{6}-16L^{r}_{8}\right]\mathcal{I}_{H,1}(\tfrac{\mathring{m}_{K}^{2}}{eH})\ ,
\label{eq:ss6}
\end{align}
which depend on the dimensionless, negative definite integrals $\mathcal{I}_{H,n}$ -- their closed form expressions are presented in \ref{app:integrals}. Both quark condensates are independent of the $\overline{\rm MS}$-bar scale, a fact that follows from the scale-invariance of the free energy, and can be verified independently using the running of the low-and-high energy constants in Eq.~\eqref{eq:Lri}, Eq.~\eqref{eq:Dri-eq7} and Eq.~\eqref{eq:Dri-eq8}.

In order to compare the light quark condensate calculated in this work to that of Ref.~\cite{ding2020chiral}, we utilize the dimensionless quantity
\begin{align}
\label{eq:qqshift}
\Sigma_{\bar{q}q}(H)&=-\frac{\hat{m}}{m_{\pi}^{2}f_{\pi}^{2}}\langle\bar{q}q\rangle_{H}+1\ ,
\end{align}
which is unity in the absence of magnetic catalysis, in which case $\langle\bar{q}q\rangle_{H}$ is zero.
The parameters used in the lattice study to generate the light quark condensate~\cite{ding2020chiral} are
\begin{align}
m_{\pi}&=220.61\ {\tt MeV}\ ,&\ m_{K}&=508.20\ {\tt MeV}\ ,&\ m_{\eta}&=684.44\ {\tt MeV}\\
\hat{m}&=9.30\ {\tt MeV}\ ,&\ m_{s}&=93.0\ {\tt MeV}\ ,&\ f_{\pi}&=96.93\ {\tt MeV}\ .
\end{align}
with the pion mass ($m_{\pi}$) and the eta mass ($m_{\eta}$) both considerably larger than their physical counterparts~\cite{Zyla:2020zbs}
\begin{align}
m_{\pi}&=139.58\ {\tt MeV}\ ,&\ m_{K}&=493.68\ {\tt MeV}\ ,&\ m_{\eta}&=547.86\ {\tt MeV}\ ,\\
\hat{m}&=3.42\ {\tt MeV}\ ,&\ \ m_{s}&=93.4\ {\tt MeV}\ ,\ &f_{\pi}&=92.21\ {\tt MeV}\ ,
\end{align}
though the kaon mass ($m_{K}$), the strange quark mass ($m_{s}$) and the pion decay constant ($f_{\pi}$) are comparable. We further require the follows $\mathcal{O}(p^{4})$~\cite{bijnens2014mesonic} and $\mathcal{O}(p^{6})$ low energy constants (LECs)~\cite{unterdorfer2008radiative}
\begin{align}
\label{eq:Lr}
10^{3}L_{4}^{r}&=0.0 \pm 0.3\ ,&\ 10^{3}L_{5}^{r}&=1.2 \pm 0.1\ ,\ &10^{3}L_{6}^{r}&=0.0 \pm 0.4\ ,\ &10^{3}L_{8}^{r}&=0.5 \pm 0.2\ ,\\
\label{eq:Cr}
10^{5}C^{r}_{61}&=1.0\pm0.3\ ,\ &10^{5}C^{r}_{62}&=0.0\pm0.2\ .
\end{align}
The physical meson masses and pion decay constant must be related to the bare quantities that appear in the condensates. Working in the isospin limit and recalling that the bare eta mass can be expressed in terms of the pion and kaon masses, we have three independent bare quantities, namely $f$ and two meson masses. Therefore, we need three physical quantities to determine these bare parameters. We can use the renormalized $m_{\pi}$, $m_{K}$, and $f_{\pi}$.  These can be calculated by utilizing the two-loop expressions for the renormalized physical masses and decay constants at least in principle though the two-loop expressions available in literature are very cumbersome. A more efficient approach is to utilize inverted expressions for the renormalized pion and kaon masses, and the pion decay constant
\begin{figure}[t]
	\centering
	\begin{subfigure}[b]{0.45\textwidth}
	\includegraphics[width=1.\textwidth]{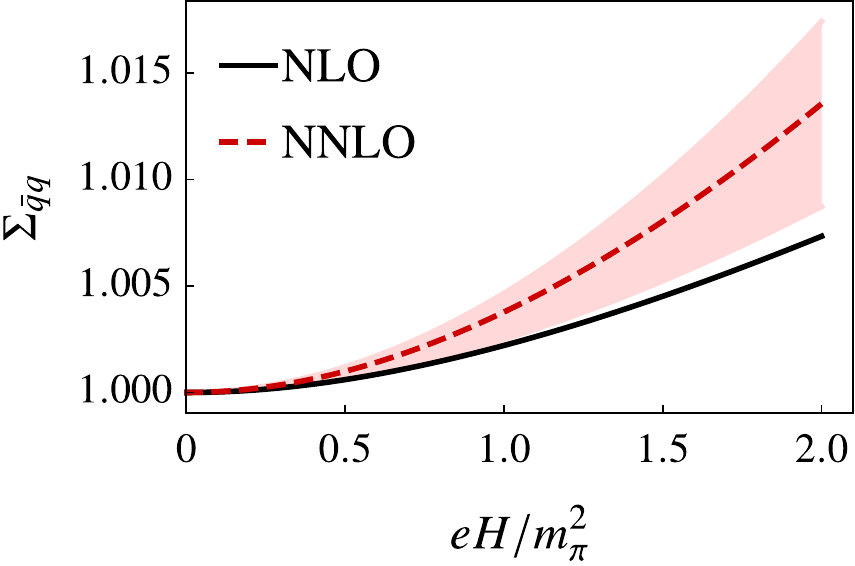}
	\end{subfigure}
	\begin{subfigure}[b]{0.45\textwidth}
	\includegraphics[width=1.\textwidth]{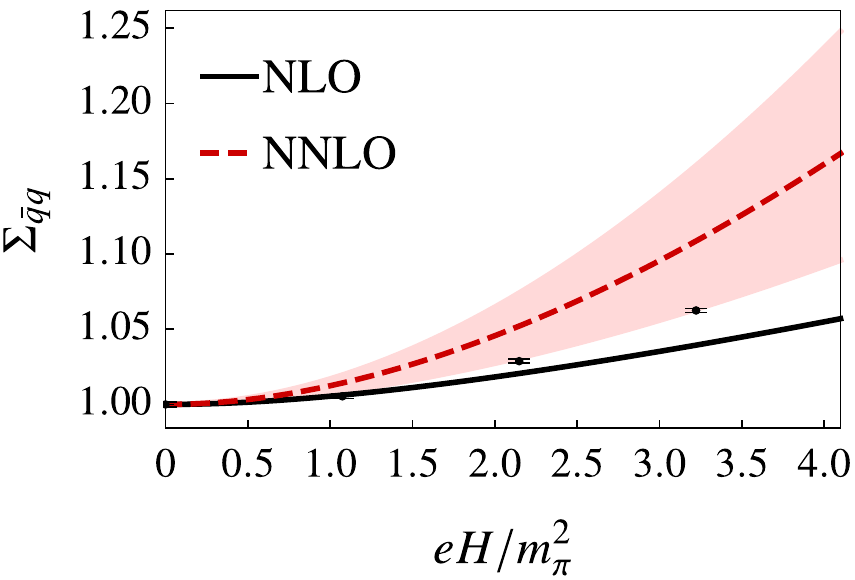}
	\end{subfigure}
\caption{Plot of the light quark condensate shift as characterized by $\Sigma_{\bar{q}q}(H)$ in Eq.~\eqref{eq:qqshift} as a function of the magnetic field ($eH$). The left panel is generated using PDG parameters~\cite{Zyla:2020zbs} while the right panel is generated using lattice parameters~\cite{ding2020chiral}.}
	\label{fig:cc}
\end{figure}
\begin{align}
\nonumber
\label{eq:pionmass3f}
\mathring{m}_{\pi}^2&=m_{\pi}^2\left[1+\left(8{L}_4^r+8{L}_5^r-16{L}_6^r-16{L}_8^r+{1\over2(4\pi)^2}\log{\Lambda^2\over m_{\pi}^2}\right){m_{\pi}^2\over f_{\pi}^{2}}+({L}_4^r-2{L}_6^r){16m_{K}^2\over f_{\pi}^{2}}\right.\\
&\left.+{m_{\eta}^2\over6(4\pi)^2f_{\pi}^2}\log{\Lambda^2\over m_{\eta}^2}\right]\ ,\\
\label{eq:kaonmass3f}
\mathring{m}_{K}^2&=m_{K}^2\left[1+\left({L}_4^r-2{L}_6^r\right){8m_{\pi}^2\over f_{\pi}^2}+(2L_4^r+{L}_5^r-4L_6^r-2{L}_8^r){8m_{\eta}^2\over f_{\pi}^2}+{m_{\eta}^2\over3(4\pi)^2f_{\pi}^2}\log{\Lambda^2\over m_{\eta}^2}\right]\ ,\\
\label{eq:piondecay3f}
f^2&=f_{\pi}^{2}\left[1-\left(8{L}_4^r+8L_5^r+{2\over(4\pi)^2}\log{\Lambda^2\over m_{\pi}^2}\right){m_{\pi}^2\over f_{\pi}^2}-\left(16L_4^r+{1\over(4\pi)^2}\log{\Lambda^2\over m_{K}^2}\right){m_{K}^2\over f_{\pi}^2}\right]\ .
\end{align}
The bare masses are then fully determined by the  $\mathcal{O}(p^{4})$ LECs.
In the $\mathcal{O}(p^{4})$ expression for the light quark condensate shift, we can replace the bare quantities with the inverted expressions above. This gives  rise to $\mathcal{O}(p^{6})$ contributions to the shift while in the $\mathcal{O}(p^{6})$ expression, we can simply replace the bare quantities with physical ones since corrections are $\mathcal{O}(p^{8})$. In the left panel of Fig.~\ref{fig:cc}, we plot $\Sigma_{\bar{q}q}$ and compare our results with the lattice while on the right panel, we plot $\Sigma_{\bar{q}q}$ using PDG parameters with the bands representing the uncertainties due to those present in the low energy constants. The monotonic increase of $\Sigma_{\bar{q}q}$ is evident in the plots. At $\mathcal{O}(p^{4})$, the lattice results agree with that of this work for $eH$ approximately equal to $m_{\pi}^{2}$ while for larger fields the $\mathcal{O}(p^{4})$ results are an underestimate. The $\mathcal{O}(p^{6})$ result, on the other hand, is consistent for all magnetic fields up to $eH\approx3.25m_{\pi}^{2}$. For the PDG parameters, the $\mathcal{O}(p^{6})$ results are consistently larger than the $\mathcal{O}(p^{4})$ but by an amount that is significantly more modest than in the right panel. For completeness, we define a quantity analogous to Eq.~\eqref{eq:qqshift} for the strange quark condensate,
\begin{align}
\label{eq:Sigmass}
\Sigma_{\bar{s}s}(H)=-\frac{\hat{m}+m_{s}}{m_{K}^{2}f_{K}^{2}}\langle \bar{s}s\rangle_{H}+1
\end{align}
that measures the magnetic catalysis associated with the strange quark condensate, which we plot in the left panel of Fig.~\ref{fig:ssmag}. As with the light quark condensate, the strange quark condensate increases monotonously with the external field, though the increase is weaker due to the significantly larger mass of the kaon. 

\subsection{Renormalized Magnetization}
The renormalized magnetization measures the first order change in the matter contribution to the free energy (density) as the external field is altered, 
\begin{align}
\label{eq:Mr}
\mathcal{M}_{r}&=-\frac{\partial\widetilde{\mathcal{F}}_{H}}{\partial (eH)}\ ,
\end{align}
where $\widetilde{\mathcal{F}}_{H}=\mathcal{F}_{H}-\frac{1}{2}H_{R}^{2}$ excludes the pure gauge contribution to the total free energy (density). Since the effect arises due to the interaction of virtual pions and kaons with the external magnetic field, the leading contribution, $\mathcal{M}_{r}^{(4)}$, appears beginning at $\mathcal{O}(p^{4})$,
\begin{align}
\mathcal{M}_{r}^{(4)}&=-\frac{1}{(4\pi)^{2}}\left[2eH\left\{\mathfrak{I}_{H}(\tfrac{\mathring{m}_{\pi}^{2}}{eH})+\mathfrak{I}_{H}(\tfrac{\mathring{m}_{K}^{2}}{eH})\right\}+(eH)^{2}\{\mathfrak{I}_{H}'(\tfrac{\mathring{m}_{\pi}^{2}}{eH})+\mathfrak{I}_{H}'(\tfrac{\mathring{m}_{K}^{2}}{eH})\}\right]\ ,
\end{align}
where prime on $\mathfrak{I}_{H}$ represents its derivative with respect to $eH$. For an alternative version of this expression, see Eq. (63) of Ref.~\cite{Adhikari:2021bou}. The next-to-leading order contribution due to two-loop diagrams is
\begin{align}\nonumber
\mathcal{M}_{r}^{(6)}=&\frac{\mathring{m}_{\pi}^4}{(4\pi f)^{2}}\left[\frac{1}{2(4\pi)^2}\log{\frac{\Lambda^{2}}{\mathring{m}_{\pi}^{2}}}+ \frac{1}{18(4\pi)^2}\log{\frac{\Lambda^{2}}{\mathring{m}_{\eta}^{2}}}+8(L_4^r+L^{r}_{5})-16(L^{r}_{6}+L^{r}_{8}) \right]\mathcal{D}'(\mathring{m}_{\pi})\\ \nonumber
+&\frac{\mathring{m}_{\pi}^{2}\mathring{m}_{K}^{2}}{(4\pi f)^{2}} \left[-\frac{1}{9(4\pi)^2}\log{\frac{\Lambda^{2}}{\mathring{m}_{\eta}^{2}}}+8(L_4^r-2L^{r}_{6})\right]\left[2\mathcal{D}'(\mathring{m}_{\pi})+\mathcal{D}'(\mathring{m}_{K})\right]\\ \nonumber
+&\frac{\mathring{m}_{K}^{4}}{(4\pi f)^{2}}\left[\frac{4}{9(4\pi)^2}\log{\frac{\Lambda^{2}}{\mathring{m}_{\eta}^{2}}}+8(2L_4^r+L^{r}_{5})-16(2L^{r}_{6}+L^{r}_{8})\right]\mathcal{D}'(\mathring{m}_{K})\\
-&\frac{4(eH)^{2}}{(4\pi f)^{2}}\left[L_9^r+L_{10}^r\right]\left[\mathcal{D}'(\mathring{m}_{\pi})+\mathcal{D}'(\mathring{m}_{K})\right]+\frac{8(eH)}{(4\pi f)^{2}}\left[L_9^r+L_{10}^r\right]\left[\mathcal{D}(\mathring{m}_{\pi})+\mathcal{D}(\mathring{m}_{K}) \right]\ ,
\label{eq:Mr}
\end{align}
where $\mathcal{D}(m_{\phi})=\frac{(eH)}{(4\pi)^{2}}\mathcal{I}_{H,2}(\tfrac{m_{\phi}^{2}}{eH})$ and $\mathcal{D}'(m_{\phi})$ is its derivative with respect to $eH$. Finally, the renormalized magnetization is scale-invariant, as follows from the scale-invariance of $\c{F}_{H}$ and $eH$.

\begin{figure}[t]
	\centering
	\begin{subfigure}[b]{0.45\textwidth}
	\includegraphics[width=1.\textwidth]{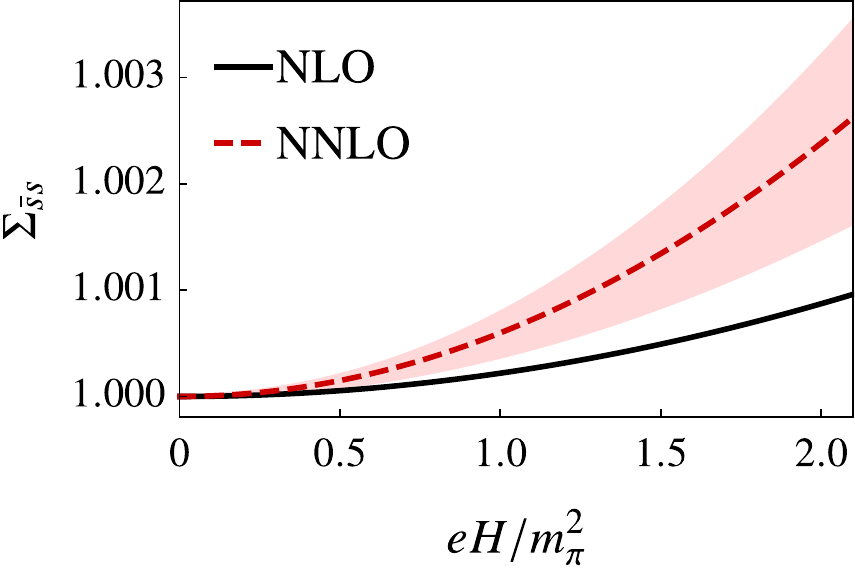}
	\end{subfigure}
	\begin{subfigure}[b]{0.45\textwidth}
	\includegraphics[width=1.\textwidth]{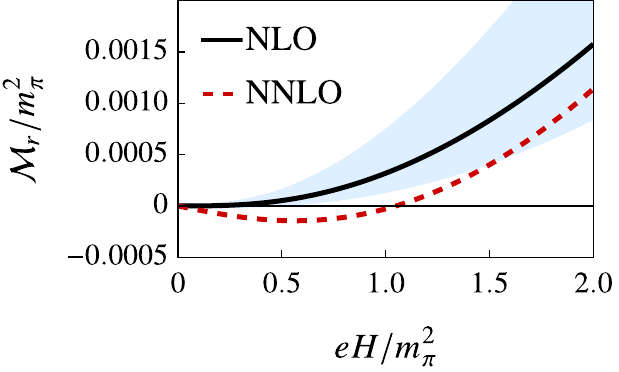}
	\end{subfigure}
\caption{Left: Plot of the shift in the strange quark condensate as characterized by $\Sigma_{\bar{s}s}$ defined in Eq.~\eqref{eq:Sigmass}. Right: Plot of the renormalized magnetization as a function of the magnetic field.}
	\label{fig:ssmag}
\end{figure}

On the right panel of Fig.~\ref{fig:ssmag}, we plot the renormalized magnetization for PDG parameters. Unlike the other plots, the size of uncertainties associated with the $\mathcal{O}(p^{6})$ LECs leads to a renormalized magnetization that almost covers the entire plot and is therefore not particularly informative. This is unlike the quark condensate shifts plotted previously. The renormalized magnetization (normalized by $m_{\pi}^{2}$) is at least an order of magnitude smaller compared to the relative shift of the quark condensates, defined in Eq.~\eqref{eq:qqshift} and \eqref{eq:Sigmass}, modulo the additive constant of plus one. Consequently, the impact of the LECs is more prominent in the plots -- this was also observed in the study of finite volume effects in a magnetic field, see Ref.~\cite{adhikari2023qcd}.  Therefore, we plot the uncertainty in the renormalized magnetization that arises due to the uncertainties in the $\mathcal{O}(p^{4})$ LECs (shown in light blue). We find that the vacuum at this order is likely to be paramagnetic, although the $\mathcal{O}(p^{6})$ result allows the possibility of a diamagnetic vacuum at low external fields. The magnetization remains negative for $eH\lesssim 0.9m_{\pi}^{2}$ and becomes positive for larger values of the magnetic field.

\section{Summary}
\label{sec:summary}
In this work, we have studied the QCD vacuum in a uniform background magnetic field using three-flavor $\chi$PT. In particular, we have calculated the vacuum free energy, light and strange quark condensate shifts and the renormalized magnetization. The calculation of the $\mathcal{O}(p^{6})$ vacuum free energy (density) is particularly non-trivial though as we have noted there are many simplifying features. By utilizing the renormalization group equations associated with the running of the low-and-high-energy constants, we have checked explicitly that the scale-dependence in the chiral logs are canceled precisely by those in the low-and-high energy constants. We also compared the light-quark condensate shift to that from a recent lattice study and find that $\mathcal{O}(p^{6})$ results are in better agreement than the result at $\mathcal{O}(p^{4})$. Finally, we studied the renormalized magnetization, which at $\mathcal{O}(p^{4})$ is positive definite but due to the uncertainties in the LECs can be either positive or negative at $\mathcal{O}(p^{6})$.

\section{Acknowledgements}
P.A. and I.S. are indebted to J.O. Andersen for collaboration on this work. The authors would also like to thank H.-T. Ding for sharing lattice data. P.A. would also like to acknowledge discussions with Brian Tiburzi and acknowledges the support of the Kavli Institute for Theoretical Physics, Santa Barbara, through which the research was supported in part by the National Science Foundation under Grant No. NSF PHY-1748958.
\appendix

\section{Useful renormalized LECs and constants for renormalization}
\label{app:usefulLEC}
In this appendix, we list quantities that are necessary for renormalizing the one and two-loop contributions to the free energy in a background magnetic field. We begin with the constants $\Gamma_{i}$ and $\Delta_{i}$ necessary to determine the running of $L^{r}_{i}$ and $H^{r}_{i}$
\begin{equation}
\begin{split}
\label{eq:GiDi}
\Gamma_{4}=\frac{1}{8},\ \Gamma_{5}=\frac{3}{8},\ \Gamma_{6}=\frac{11}{144},\ \Gamma_{7}=0,\ \Gamma_{8}=\frac{5}{48},\ \Gamma_{9}=\frac{1}{4},\ \Gamma_{10}=-\frac{1}{4},\ \Delta_{1}=-\frac{1}{8},\ \Delta_{2}=\frac{5}{24}\ .
\end{split}
\end{equation} 
The running of the renormalized low-energy-constant that appears in the $\mathcal{O}(p^{6})$ Lagrangian requires the constants $\Gamma_{19}^{(2)}$ and $\Gamma_{19}^{(1)}$ listed below
\begin{align}
\label{eq:G2i}
&\Gamma_{19}^{(2)}=\frac{11}{1944},\ \Gamma_{20}^{(2)}=\frac{13}{1296},\ \Gamma_{21}^{(2)}=\frac{59}{3888},\ \Gamma_{61}^{(2)}=0,\ \Gamma_{62}^{(2)}=0,\ \Gamma^{(2)}_{94}=-\frac{119}{162}\\
\label{eq:G1i}
&\Gamma_{19}^{(1)}=\frac{1}{(4\pi)^{2}}\frac{1517}{93312},\ \Gamma_{20}^{(1)}=-\frac{1}{(4\pi)^{2}}\frac{1517}{62208},\ \Gamma_{21}^{(1)}=\frac{1}{(4\pi)^{2}}\frac{1517}{186624},\ \Gamma_{61}^{(1)}=0\\
\label{eq:G1imore}
&\Gamma_{62}^{(1)}=0,\ \Gamma^{(1)}_{94}=-\frac{1}{(4\pi)^{2}}\frac{1517}{7776}\ .
\end{align}
$\Gamma^{(L)}_{i}$ are particular linear combinations of the $\mathcal{O}(p^{4}$ LECs that determine the running of $D^{r}_{i}$,
\begin{align}
\label{eq:GammaL19}
&\Gamma^{(L)}_{19}=+\frac{44}{81}L_{1}^{r}+\frac{8}{81}L_{2}^{r}+\frac{10}{81}L_{3}^{r}+\frac{1}{3}L_{4}^{r}-\frac{1}{4}L_{5}^{r}+\frac{4}{27}L_{6}^{r}-\frac{32}{27}L_{7}^{r}-\frac{1}{6}L_{8}^{r}\\
\label{eq:GammaL20}
&\Gamma^{(L)}_{20}=-\frac{22}{27}L_{1}^{r}-\frac{4}{27}L_{2}^{r}-\frac{5}{27}L_{3}^{r}+\frac{1}{3}L_{4}^{r}+\frac{13}{12}L_{5}^{r}-\frac{17}{9}L_{6}^{r}+\frac{1}{9}L_{7}^{r}-\frac{31}{18}L_{8}^{r}\\
\label{eq:GammaL21}
&\Gamma^{(L)}_{21}=+\frac{22}{81}L_{1}^{r}+\frac{4}{81}L_{2}^{r}+\frac{5}{81}L_{3}^{r}+\frac{7}{9}L_{4}^{r}-\frac{1}{27}L_{5}^{r}-\frac{31}{27}L_{6}^{r}+\frac{5}{27}L_{7}^{r}\\
\label{eq:GammaL61}
&\Gamma^{(L)}_{61}=-\frac{3}{4}(L_{9}^{r}+L^{r}_{10})\\
\label{eq:GammaL62}
&\Gamma^{(L)}_{62}=-\frac{1}{4}(L_{9}^{r}+L^{r}_{10})\\
\label{eq:GammaL94}
&\Gamma^{(L)}_{94}=-\frac{176}{27}L_{1}^{r}-\frac{32}{27}L_{2}^{r}-\frac{40}{27}L_{3}^{r}-4L^{r}_{4}-\frac{16}{9}L_{6}^{r}-\frac{88}{9}L^{r}_{7}\ ,
\end{align}
which are useful to determine their running. For succintness, we only list the running of the particular combinations of $D^{r}_{i}$ that appear in the free energy, chiral condensates and renormalized magnetization,
\begin{align}
\label{eq:Dri-eq1}
\Lambda\frac{d}{d\Lambda}[D^{r}_{19}+D^{r}_{20}+D^{r}_{21}]&=\frac{13}{9}L_{4}^{r}+\frac{43}{54}L_{5}^{r}-\frac{26}{9}L_{6}^{r}-\frac{8}{9}L_{7}^{r}-\frac{17}{9}L_{8}^{r}\\
\label{eq:Dri-eq2}
\Lambda\frac{d}{d\Lambda}[3D^{r}_{19}+D^{r}_{20}-D^{r}_{21}]&=-L_{4}^{r}+\frac{4}{9}L_{5}^{r}+2L_{6}^{r}-4L_{7}^{r}-\frac{20}{9}L_{8}^{r}\\
\label{eq:Dri-eq3}
\Lambda\frac{d}{d\Lambda}[12D^{r}_{19}+4D^{r}_{20}+12D^{r}_{21}+D^{r}_{94}]&=\frac{32}{3}L_{4}^{r}+\frac{8}{9}L_{5}^{r}-\frac{64}{3}L_{6}^{r}-\frac{64}{3}L_{7}^{r}-\frac{80}{9}L_{8}^{r}\\
\label{eq:Dri-eq4}
\Lambda\frac{d}{d\Lambda}[4D^{r}_{19}+12D^{r}_{20}+4D^{r}_{21}-D^{r}_{94}]&=\frac{112}{9}L_{4}^{r}+\frac{320}{27}L_{5}^{r}-\frac{224}{9}L_{6}^{r}+\frac{64}{9}L_{7}^{r}-\frac{64}{3}L_{8}^{r}\\
\label{eq:Dri-eq5}
\Lambda\frac{d}{d\Lambda}[2D^{r}_{61}+3D^{r}_{62}]&=-\frac{9}{4}(L_{9}^{r}+L_{10}^{r})\\
\label{eq:Dri-eq6}
\Lambda\frac{d}{d\Lambda}[D^{r}_{61}+6D^{r}_{62}]&=-\frac{9}{4}(L_{9}^{r}+L_{10}^{r})\\
\label{eq:Dri-eq7}
\Lambda\frac{d}{d\Lambda}\left[\frac{160}{9}D^{r}_{61}+\frac{128}{3}D^{r}_{62}\right]&=-24(L_{9}^{r}+L_{10}^{r})\\
\label{eq:Dri-eq8}
\Lambda\frac{d}{d\Lambda}\left[\frac{32}{9}D^{r}_{61}+\frac{64}{3}D^{r}_{62}\right]&=-8(L_{9}^{r}+L_{10}^{r})\ .
\end{align}

\section{$\chi$PT Lagrangian required to compute the $\mathcal{O}(p^{6})$ free energy}
\label{app:L}
The free energy calculation requires the four-meson contribution from the $\c{O}(p^{2})$ Lagrangian 
\begin{equation}
\begin{split}
\mathcal{L}_{2,4}&=\frac{1}{24f^{2}}\mathring{m}^{2}_{\pi}(\pi^{0})^{4}+\frac{1}{12f^{2}}\mathring{m}^{2}_{\pi}(\pi^{0})^{2}\eta^{2}+\frac{1}{6f^{2}}\mathring{m}^{2}_{\pi}\pi^{+}\pi^{-}\eta^{2}+\frac{1}{216f^{2}}(16\mathring{m}^{2}_{K}-7\mathring{m}^{2}_{\pi})\eta^{4}\\
&-\frac{1}{6f^{2}}[2(\pi^{0})^{2}D_{\mu}\pi^{+}D^{\mu}\pi^{-}+\pi^{+}\pi^{-}\{2\partial_{\mu}\pi^{0}\partial^{\mu}\pi^{0}-\mathring{m}_{\pi}^{2}(\pi^{0})^{2}\}]\\
&-\frac{1}{6f^{2}}\pi^{+}\pi^{-}[2D_{\mu}\pi^{+}D^{\mu}\pi^{-}-\mathring{m}_{\pi}^{2}\pi^{+}\pi^{-}+D_{\mu}K^{+}D^{\mu}K^{-}-\mathring{m}_{K}^{2}K^{+}K^{-}]\\
&-\frac{1}{6f^{2}}K^{+}K^{-}[2D_{\mu}K^{+}D^{\mu}K^{-}-\mathring{m}_{K}^{2}K^{+}K^{-}+D_{\mu}\pi^{+}D^{\mu}\pi^{-}-\mathring{m}_{\pi}^{2}\pi^{+}\pi^{-}]\\
&-\frac{1}{6f^{2}}[2K^{0}\bar{K}^{0}\partial_{\mu}K^{0}\partial^{\mu}\bar{K}^{0}-\mathring{m}_{K}^{2}(K^{0}\bar{K}^{0})^{2}]\\
&-\frac{1}{12f^{2}}[K^{0}\bar{K}^{0}\partial_{\mu}\pi^{0}\partial^{\mu}\pi^{0}+(\pi^{0})^{2}\partial_{\mu}K^{0}\partial^{\mu}\bar{K}^{0}-(\mathring{m}_{\pi}^{2}+\mathring{m}_{K}^{2})(\pi^{0})^{2}K^{0}\bar{K}^{0}]\\
&-\frac{1}{12f^{2}}[K^{+}K^{-}\partial_{\mu}\pi^{0}\partial^{\mu}\pi^{0}+(\pi^{0})^{2}D_{\mu}K^{+}D^{\mu}K^{-}-(\mathring{m}_{\pi}^{2}+\mathring{m}_{K}^{2})(\pi^{0})^{2}K^{+}K^{-}]\\
&-\frac{1}{12f^{2}}[K^{0}\bar{K}^{0}D_{\mu}\pi^{+}D^{\mu}\pi^{-}+\pi^{+}\pi^{-}\partial_{\mu}K^{0}\partial^{\mu}\bar{K}^{0}-(\mathring{m}_{\pi}^{2}+\mathring{m}_{K}^{2})\pi^{+}\pi^{-}K^{0}\bar{K}^{0}]\\
&-\frac{1}{6f^{2}}[K^{+}K^{-}\partial_{\mu}K^{0}\partial^{\mu}\bar{K}^{0}+K^{0}\bar{K}^{0}D_{\mu}K^{+}D^{\mu}K^{-}-2\mathring{m}_{K}^{2}K^{+}K^{-}K^{0}\bar{K}^{0}]\\
&-\frac{1}{12f^{2}}[3K^{+}K^{-}\partial_{\mu}\eta\partial^{\mu}\eta+3\eta^{2}D_{\mu}K^{+}D^{\mu}K^{-}+(\mathring{m}_{\pi}^{2}-3\mathring{m}_{K}^{2})K^{+}K^{-}\eta^{2}]\\
&-\frac{1}{12f^{2}}[3K^{0}\bar{K}^{0}\partial_{\mu}\eta\partial^{\mu}\eta+3\eta^{2}\partial_{\mu}K^{0}\partial^{\mu}\bar{K}^{0}+(\mathring{m}_{\pi}^{2}-3\mathring{m}_{K}^{2})K^{0}\bar{K}^{0}\eta^{2}]\ ,
\end{split}
\end{equation}
the tree-level contribution from the $\mathcal{O}(p^{4})$ Lagrangian
\begin{equation}
\begin{split}
\mathcal{L}_{4,0}&=(4L_{6}-2L_{8}-H_{2})(\mathring{m}^{2}_{\pi}+2\mathring{m}_{K}^{2})^{2}+4(2L_{8}+H_{2})(\mathring{m}_{\pi}^{4}+2\mathring{m}_{K}^{4})+\frac{4}{3}(L_{10}+2H_{1})(eH)^{2}\ .
\end{split}
\end{equation}
the two-meson contribution from the $\c{O}(p^{4})$ Lagrangian
\begin{align}
\nonumber
\mathcal{L}_{4,2}&=\frac{4L_{4}}{f^{2}}(\mathring{m}_{\pi}^{2}+2\mathring{m}_{K}^{2})[2D_{\mu}\pi^{+}D^{\mu}\pi^{-}+\partial_{\mu}\pi^{0}\partial^{\mu}\pi^{0}+2D_{\mu}K^{+}D^{\mu}K^{-}+2\partial_{\mu}K^{0}\partial^{\mu}\bar{K}^{0}+\partial_{\mu}\eta\partial^{\mu}\eta]\\ \nonumber
&+\frac{4L_{5}}{f^{2}}[\mathring{m}_{\pi}^{2}(2D_{\mu}\pi^{+}D^{\mu}\pi^{-}+\partial_{\mu}\pi^{0}\partial^{\mu}\pi^{0})+2\mathring{m}_{K}^{2}(D_{\mu}K^{+}D^{\mu}K^{-}+\partial_{\mu}K^{0}\partial^{\mu}\bar{K}^{0})+\mathring{m}_{\eta}^{2}\partial_{\mu}\eta\partial^{\mu}\eta]\\ \nonumber
&-\frac{8L_{6}}{f^{2}}(\mathring{m}_{\pi}^{2}+2\mathring{m}_{K}^{2})[\mathring{m}_{\pi}^{2}\{2\pi^{+}\pi^{-}+(\pi^{0})^{2}\}+2\mathring{m}_{K}^{2}(K^{+}K^{-}+K^{0}\bar{K}^{0})+\mathring{m}_{\eta}^{2}\eta^{2}]\\ \nonumber
&-\frac{64L_{7}}{3f^{2}}(\mathring{m}_{\pi}^{2}-\mathring{m}_{K}^{2})^{2}\eta^{2}\\ \nonumber
&-\frac{16L_{8}}{f^{2}}[\mathring{m}_{\pi}^{4}\pi^{+}\pi^{-}+\tfrac{1}{2}\mathring{m}_{\pi}^{4}(\pi^{0})^{2}+\mathring{m}_{K}^{4}(K^{+}K^{-}+K^{0}\bar{K}^{0})+\tfrac{1}{3}(4\mathring{m}_{K}^{4}-4\mathring{m}_{\pi}^{2}\mathring{m}_{K}^{2}+\tfrac{3}{2}\mathring{m}_{\pi}^{4})\eta^{2}]\\ 
&+\frac{2iL_{9}}{f^{2}}eF_{\mu\nu}(D^{\mu}\pi^{+}D^{\nu}\pi^{-}+D^{\mu}K^{+}D^{\nu}K^{-})-\frac{2L_{10}}{f^{2}}(eF_{\mu\nu})(eF^{\mu\nu})(\pi^{+}\pi^{-}+K^{+}K^{-})\ ,
\end{align}
and the tree-level contribution from the $\mathcal{O}(p^{6})$ Lagrangian
\begin{align}
\nonumber
\mathcal{L}_{6,0}&=2\mathring{m}_{\pi}^{6}[4C_{19}+12C_{20}+4C_{21}-C_{94}]+4\mathring{m}_{\pi}^{4}\mathring{m}_{K}^{2}[12C_{19}+4C_{20}+12C_{21}+C_{94}]\\ \nonumber
&-32\mathring{m}_{\pi}^{2}\mathring{m}_{K}^{4}[3C_{19}+C_{20}-3C_{21}]+64\mathring{m}_{K}^{6}[C_{19}+C_{20}+C_{21}]\\ 
&+\frac{32}{9}(eH)^{2}\mathring{m}_{\pi}^{2}[2C_{61}+3C_{62}]+\frac{32}{9}(eH)^{2}\mathring{m}_{K}^{2}[C_{61}+6C_{62}]\ .
\end{align}

\section{Useful Integrals}
\label{app:integrals}
The contribution to the one-loop effective potential of a pair of charge mesons, $\phi$ and $\phi^{\dagger}$ in Eq.~(\ref{eq:IH}) is
\begin{equation}
\begin{split}
I_{H}(m_{\phi})&=\frac{eH}{2\pi}\sum_{N=0}^{\infty}\int_{p_{0},p_{z}}\ln[p_{0}^{2}+p_{z}^{2}+m_{H}^{2}]\ ,
\end{split}
\end{equation}
where the sum is over the Landau levels, $m_{H}^{2}=m_{\phi}^{2}+(2N+1)|eH|$ and the integrals are $\int_{p_{0}p_{z}}\equiv\int\frac{dp_{0}}{2\pi}\frac{dp_{z}}{2\pi}$. The divergence is independent of the magnetic field,
\begin{align}
I_{H}(m_{\phi})=&I_{H}^{\rm div}(m_{\phi})+I_{H}^{\rm fin}(m_{\phi})\\
I_{H}^{\rm div}(m_{\phi})=&-\frac{m_{\phi}^{4}}{2(4\pi)^{2}}\left[\frac{1}{\epsilon}+\frac{3}{2}+\log\frac{\Lambda^{2}}{m_{\phi}^{2}} \right]+\frac{(eH)^{2}}{6(4\pi)^{2}}\left[\frac{1}{\epsilon}+\log\frac{\Lambda^{2}}{m_{\phi}^{2}}\right]\\
I_{H}^{\rm fin}(m_{\phi})=&-\frac{1}{(4\pi)^{2}}\int_{0}^{\infty} \frac{ds}{s^{3}}e^{-m_{\phi}^{2}s}\left[\frac{eHs}{\sinh eHs}-1+\frac{(eHs)^{2}}{6}\right]\ .
\end{align}
The finite contribution can be written in terms of a dimensionless integral, $\mathfrak{I}_{H}(z)$, where $z$ is a dimensionless ratio, $z=\frac{m_{\phi}^{2}}{eH}$,
\begin{equation}
\begin{split}
\label{eq:Schwinger-integral}
I_{H}^{\rm fin}(m_{\phi})&=\frac{(eH)^{2}}{(4\pi)^{2}}\mathfrak{I}_{H}(z)\\
\mathfrak{I}_{H}(z)&=-\int_{0}^{\infty}dy\ \frac{e^{-zy}}{y^{3}}\left[\frac{y}{\sinh y}-1+\frac{y^{2}}{6}\right]\\
&=4\zeta^{(1,0)}(-1,\tfrac{z+1}{2})+(\tfrac{z}{2})^{2}(1-2\log \tfrac{z}{2})+\tfrac{1}{6}(\log \tfrac{z}{2}+1)\ ,
\end{split}
\end{equation}
 where $\zeta(s,a)$ is the Hurwitz zeta function with the two numbers in the subscripts indicating the number of derivatives with respect to $s$ and $a$ respectively.

The coincident Schwinger propagator of Eq.~(\ref{eq:D}) can be written in terms of the dimensionless, negative definite integral
\begin{equation}
\begin{split}
\label{eq:IHn}
\mathcal{I}_{H,n}(z)&=\int_{0}^{\infty}dy\frac{e^{-zy}}{y^{n}}\left(\frac{y}{\sinh y}-1\right)\ .
\end{split}
\end{equation}
We only require the $n=2$ and $n=1$ integrals presented below, see Eq.~(\ref{eq:IH2GR}) for an alternate version of the former.
\begin{align}
\label{eq:IH2}
\mathcal{I}_{H,2}(z)&=2\zeta^{(1,0)}(0,\tfrac{z+1}{2})-z(\log \tfrac{z}{2}-1)\\
\label{eq:IH1}
\mathcal{I}_{H,1}(z)&=\log\tfrac{z}{2}-\psi_{0}\left (\tfrac{z+1}{2} \right)\ .
\end{align}
$\psi_{n}(z)$ is the polygamma function that is related to the $\Gamma(z)$ function through derivatives
\begin{align}
\psi_{n}(z)=\frac{d^{n+1}}{dz^{n+1}}\log\Gamma(z)\ .
\end{align}

\section{Vacuum Free Energy (Density)}
\label{app:freeenergy}
The vacuum free energy (density) in the absence of the external field is 
\begin{align}
\mathcal{F}_{0}=\mathcal{F}_{0}^{(2)}+\mathcal{F}_{0}^{(4)}+\mathcal{F}_{0}^{(6)}\ ,
\end{align}
where $\mathcal{F}_{0}^{(n)}$ is the $\mathcal{O}(p^{n})$ contribution, 
\begin{align} 
\mathcal{F}^{(2)}_{0}&=\frac{f^{2}}{2}(\mathring{m}_{\pi}^{2}+2\mathring{m}_{K}^{2})\\ \nonumber
\mathcal{F}_{0}^{(4)}&=-(4L^{r}_{6}-2L^{r}_{8}-H^{r}_{2})(\mathring{m}_{\pi}^{2}+2\mathring{m}_{K}^{2})^{2}-4(2L^{r}_{8}+H^{r}_{2})(\mathring{m}_{\pi}^{4}+2\mathring{m}_{K}^{4})\\
&-\frac{3\mathring{m}_{\pi}^{4}}{4(4\pi)^{2}}\left[\frac{1}{2}+\log\frac{\Lambda^{2}}{\mathring{m}_{\pi}^{2}}\right]-\frac{\mathring{m}_{\pi}^{4}}{(4\pi)^{2}}\left[\frac{1}{2}+\log\frac{\Lambda^{2}}{\mathring{m}_{K}^{2}}\right]-\frac{\mathring{m}_{\pi}^{4}}{4(4\pi)^{2}}\left[\frac{1}{2}+\log\frac{\Lambda^{2}}{\mathring{m}_{\eta}^{2}}\right]
\end{align}
\begin{align}
\nonumber
\mathcal{F}_{0}^{(6)}&=\frac{\mathring{m}_{\pi}^{6}}{(4\pi f)^{2}}\Bigg[-8D_{19}^{r}-24D_{20}^{r}-8D_{21}^{r}+2D_{94}^{r}+12\left(L_{4}^{r}+L_{5}^{r}-2L_{6}^{r}-2L_{8}^{r}\right)\log\frac{\Lambda^2}{\mathring{m}_{\pi}^{2}}\Bigg.\\ \nonumber
&\Bigg.+\frac{4}{27}\left(3L^{r}_{4}-L^{r}_{5}-6L^{r}_{6}+48L^{r}_{7}+18L^{r}_{8}\right)\log\frac{\Lambda^2}{\mathring{m}_{\eta}^2}+\frac{3}{8(4\pi)^2}\left(\log\frac{\Lambda^2}{\mathring{m}_{\pi}^2}\right)^2\Bigg.\\ \nonumber
&\Bigg.+\frac{1}{12(4\pi)^{2}}\log\frac{\Lambda^2}{\mathring{m}_{\pi}^2}\log\frac{\Lambda^2}{\mathring{m}_{\eta}^2}+\frac{7}{648(4\pi)^{2}}\left(\log\frac{\Lambda^2}{\mathring{m}_{\eta}^2}\right)^2\Bigg]\\ \nonumber
&+\frac{\mathring{m}_{\pi}^{4}\mathring{m}_{K}^{2}}{(4\pi f)^{2}}\Bigg[-48D_{19}^{r}-16D_{20}^{r}-48D_{21}^{r}-4D_{94}^{r}+24\left(L_{4}^{r}-2L_{6}^{r}\right)\log\frac{\Lambda^2}{\mathring{m}_{\pi}^{2}}\Bigg.\\ \nonumber
&\Bigg.-\frac{8}{9}\left(3L_{4}^{r}-2L_{5}^{r}-6L_{6}^{r}+48L_{7}^{r}+20L_{8}^{r}\right)\log\frac{\Lambda^2}{\mathring{m}_{\eta}^2}\Bigg.\\ \nonumber
&\Bigg.-\frac{1}{3(4\pi)^{2}}\log\frac{\Lambda^2}{\mathring{m}_{\pi}^2}\log\frac{\Lambda^2}{\mathring{m}_{\eta}^2}-\frac{1}{9(4\pi)^{2}}\left(\log\frac{\Lambda^2}{\mathring{m}_{\eta}^2}\right)^2\Bigg]\\ \nonumber
&+\frac{\mathring{m}_{\pi}^{2}\mathring{m}_{K}^{4}}{(4\pi f)^{2}}\Bigg[96D_{19}^{r}+32D_{20}^{r}-96D_{21}^{r}+16\left(L_{4}^{r}-2L_{6}^{r}\right)\log\frac{\Lambda^2}{\mathring{m}_{K}^{2}}\Bigg.\\ \nonumber
&\Bigg.-\frac{2}{9}\left(32L_{5}^{r}-288L_{7}^{r}-160L_{8}^{r}\right)\log\frac{\Lambda^2}{\mathring{m}_{\eta}^2}-\frac{2}{9(4\pi)^2}\log\frac{\Lambda^2}{\mathring{m}_{K}^2}\log\frac{\Lambda^2}{\mathring{m}_{\eta}^2}+\frac{10}{27(4\pi)^{2}}\left(\log\frac{\Lambda^2}{\mathring{m}_{\eta}^2}\right)^2\Bigg]\\ \nonumber
&+\frac{\mathring{m}_{K}^{6}}{(4\pi f)^{2}}\Bigg[-64D_{19}^{r}-64D_{20}^{r}-64D_{21}^{r}+16\left(2L_{4}^{r}+L_{5}^{r}-4L_{6}^{r}-2L_{8}^{r}\right)\log\frac{\Lambda^2}{\mathring{m}_{K}^{2}}\Bigg.\\ \nonumber
&\Bigg.+\frac{128}{27}\left(3L^{r}_{4}+2L_{5}^{r}-6L^{r}_{6}-6L^{r}_{7}-6L^{r}_{8}\right)\log\frac{\Lambda^2}{\mathring{m}_{\eta}^2}\Bigg.\\
&\Bigg.+\frac{8}{9(4\pi)^2}\log\frac{\Lambda^2}{\mathring{m}_{K}^2}\log\frac{\Lambda^2}{\mathring{m}_{\eta}^2}-\frac{32}{81(4\pi)^{2}}\left(\log\frac{\Lambda^2}{\mathring{m}_{\eta}^2}\right)^2\Bigg]\ .
\end{align}

\bibliographystyle{apsrev4-1}
\bibliography{bib}
\end{document}